\documentclass[pdftex,12pt]{article}
\usepackage{amsmath,amsfonts,amsbsy,latexsym,amssymb,amscd,amstext}
\usepackage{pst-node}
\usepackage{dsfont}
\usepackage{overpic,youngtab}
\pdfoutput=1
\usepackage[font=small,labelfont=bf]{caption}
\usepackage{rotating}
\usepackage{tikz}
\usepackage{slashed}
\usepackage{graphicx}

\def\be{\begin{equation}}
\def\ee{\end{equation}}
\def\bea{\begin{eqnarray}}
\def\eea{\end{eqnarray}}
\def\pd{\partial}
\def\a{\alpha}
\def\b{\beta}
\def\xp{\x^{\prime}}
\def\vxp{\vec{x}^\prime}
\def\vx{\vec{x}}

\def\g{\gamma}
\def\d{\delta}
\def\m{\mu}

\def\n{\nu}
\def\t{\tau}

\def\l{\lambda}
\def\cR{\cal R}
\def\lp{\lambda^\prime}

\def\r{\rho}

\def\vx{\vec{x}}

\def\cR{\cal{R}}

\def\bg{\bar{g}}

\def\xp{x^\prime}

\def\bp{\bar{\phi}}
\def\s{\sigma}
\def\e{\epsilon}

\def\bma{\begin{pmatrix}}
\def\ema{\end{pmatrix}}

\def\bp{\bar{\phi}}

\def\bg{\bar{g}}
\def\bi{\begin{itemize}}
\def\ei{\end{itemize}}
\def\bn{\bar{\nabla}}

\def\bp{\bar{\phi}}

\begin{document}

		\vspace*{-1cm}
		\phantom{hep-ph/***} 
		{\flushleft
			{{FTUAM-20-8}}
			\hfill{{ IFT-UAM/CSIC-20-58}}}
		\vskip 1.5cm
		\begin{center}
		{\LARGE\bfseries Windows on quantum gravity.}\\[3mm]
			\vskip .3cm
		
		\end{center}
		\vskip 0.5  cm
		\begin{center}
			{\large Enrique \'Alvarez }
			\\
			\vskip .7cm
			{
				Departamento de F\'isica Te\'orica and Instituto de F\'{\i}sica Te\'orica, 
				IFT-UAM/CSIC,\\
				Universidad Aut\'onoma de Madrid, Cantoblanco, 28049, Madrid, Spain\\
				\vskip .1cm

				\vskip .5cm
				\begin{minipage}[l]{.9\textwidth}
					\begin{center} 
						\textit{E-mail:} 
						{enrique.alvarez@uam.es},
					
					\end{center}
				\end{minipage}
			}
		\end{center}
	\thispagestyle{empty}
	
\begin{abstract}
	\noindent
These are notes for a graduate course in quantum gravity planned  at IFT-UAM/CSIC for the spring term of 2020, but  delayed for  an indefinite period of time (although a shortened version was given in the  Escuela de F\'{\i}sica (Institute of Physics) in the University of Costa Rica.). The aim of the course was to highlight the most important conceptual problems in the field and to summarize some of the most imaginative solutions to them that have been proposed in the literature. The course can best be characterized as  idiosyncratic rather than  encyclopedic. 
\end{abstract}

\newpage
\tableofcontents
	\thispagestyle{empty}
\flushbottom

\thispagestyle{empty}
\newpage
\setcounter{page}{1} 

\section{Introduction.}
 The aim of these notes is to assess progress in the field in the last few decades (since the advent of strings, say). I have tried  to avoid unnecessary technicalities, to the extent that this is compatible with making precise statements. Previous references written much in the same spirit as this one are \cite{Weinberg,Alvarez,AlvarezGaume,GraciaBondia}
\par 
\par
It can not be denied that progress in this particular field has been largely overemphasized.
\par
The problem of quantum gravity is not by any means a new one. It is not even clear to what extent it is a scientific  problem. Certainly, there is not a single experiment that points out where we have to look. In spite of this, the list of people that have attempted to understand something about the relationship between General Relativity (GR) and Quantum Mechanics (QM) is quite large; there are books devoted entirely to it \cite{Books}.
\par
Attempts  can be roughly classified in two big groups after some coarse graining. The staring point is  that we have two theories which have experimental confirmation in their respective regions of validity and up to the often astounding experimental precision, namely General Relativity(GR), valid when $\hbar=0$ (but $G\neq 0$ and $c\leq \infty$)  and quantum mechanics (QM), valid when $\hbar\neq 0$ but $G=0$ and $c\leq \infty$. The region with $G=0$ but $c\leq \infty$ is also understood in some sense; it corresponds to quantum field theory (QFT).
\par
Then it appears natural to some people to put General Relativity (GR) first, and try in some sense to quantize the space-time geometry.
\par
 On the other side of the coin there are those that put Quantum Mechanics (QM) first, and try to change the gravitational theories accordingly. I include in this group theories in which GR is treated as an emergent theory, and the fundamental quantum variables are not necessarily geometrical.
\par
Another contentious point is whether it should be demanded that the ensuing hypothetical theory  obeys a {\em correspondence principle} of sorts. By this we mean that whatever theory is put forward, one of the essential requisites is that it  should be smoothly connected with Quantum Field Theory (QFT) in some smooth limit and with GR in an also smooth way.
\par
As I said, not everybody shares this principle (I do).
\par
 The aim of these notes has certainly not been to give a complete review of the field. I only touch a few topics which seem to me particularly interesting, and about which I think I am able to make some comments. Also the depth on the treatment is quite uneven. This is sometimes due to my taste, but also sometimes the reason is purely pedagogical. There are some things (like the canonical approach) that in my opinion every serious student should master. Advance apologies to  all authors whose works  are not duly represented. This is most probably due to ignorance for my part.
 \par
 The first paragraph is devoted to a short discussion of whatever experimental hard evidence we have (or may be we can have in the future) that gravity has necessarily to be quantized.
 \par
 Then some of the general arguments on how Einstein's equations can emerge from a more fundamental theory are reviewed.
 \par
 After that  some thought is devoted to what are the questions we would like to have our theory to answer, and we find that even to make a precise statement of those questions is  not an easy task.
 \par
  Then we devote some space to a few  comments about the canonical formalism, which is always at the root of any quantum mechanical approach. This is probably the longest, and in my opinion the most important one.
  \par
  
 It follows some thoughts  on what is the symmetry group of the theory and the related question of what are the observables.
 \par
 We end we some short comments on what has been the impact of string theory on the topics above.

\section{Do we need quantum gravity?}
The first thing we have to decide is whether there is any evidence that gravity has to be quantized. ({\em confer} references and comments in \cite{Carlip}).
If gravity remains a classical field,  the second member of Einstein's equations should be interpreted as some expectation value of a composite operator in QFT.
\be
T_{\m\n}\rightarrow \langle \psi\left|T_{\m\n}\right|\psi\rangle
\ee
This equation (apparently first proposed by M\o ller \cite{Moller}) is not easy to justify from first principles. In spite of this fact, it is frequently used to compute the {\em gravitational back-reaction} to some quantum corrections to some particular phenomenon under consideration, computed in some initially fixed gravitational background.
\par
 It has been argued \cite{Duff}\cite{Anselmi}  that these modified Einstein's equations are  not invariant under field redefinitions. Besides, renormalization forces to include terms quadratic in curvature that in turn are claimed to imply negative energy \cite{Woodard}.
\par
 Perhaps the most precise analysis of M\o ller's ansatz is the one of Randjbar-Daemi and Kibble \cite {Kibble} where they deduce it from the variational principle.
\be
S_\psi\equiv {1\over 2 \kappa^2}\int d(vol)R+\int dt \bigg\{\text{Im}\langle\dot{\psi}|\psi\rangle-\langle \psi \left|H\right|\psi\rangle+\a\left(\langle\psi|\psi\rangle-1\right)\bigg\}
\ee
The equations of motion (EM) read
\bea
&i{\pd\over \pd t}|\psi\rangle=H\,|\psi\rangle-\a|\psi\rangle\nonumber\\
&\langle\psi|\psi\rangle-1=0\nonumber\\
&G_{\m\n}= 2\kappa^2 \left\langle \psi\left|{\d H\over \d g^{\m\n}}\right|\psi\right\rangle
\eea
where 
\be
G_{\m\n}\equiv R_{\m\n}-{1\over 2} R g_{\m\n}
\ee
is Einstein's tensor.
\par
Here the hamiltonian $H$ depends on the gravitational field, which depends on the state $|\psi\rangle$. This then introduces non-linerities in quantum mechanics.
Those se are severely restricted by experiment \cite{Weinbergqm} to be at most of the order of $10^{-21}$. Whether this is in contradiction with the semiclassical theory is not clear \cite{Carlip}.
\par
 A quick (newtonian) estimate can be done as follows. Assume the equations
 \bea
 &i\hbar{\pd\psi\over \pd t}=-{\hbar^2\over 2 m} \Delta\,\psi-m \Phi \psi\nonumber\\
 &\Delta \, \Phi= 4 \pi G m |\psi|^2
 \eea
Consider  a particle of mass $m$ with a localized initial wavefunction
\be
\psi(r,t=0)\equiv \left({1\over \pi \Delta x^2}\right)^{3\over 4}\,e^{-{r^2\over 2 \Delta x^2}}
\ee
The time evolution is driven by two effects. The ordinary spread of the wave function in QM on the one hand, and the self-gravitation on the other. Which of those effects dominates depends on the particle mass.
The peak probability density for a free particle is known to occur at
\be
{r_p \over\Delta x}\sim \sqrt{1+{\hbar^2\over  m^2 \Delta x^4}t^2}
\ee
This accelerates outward as
\be
\ddot{r}_p\sim {\hbar^2\over m^2 r_p^3}
\ee
The opposite inward gravitational acceleration is
\be
a_g\sim {G m\over r_p^2}
\ee
Equality at t=0 needs
\be
m_c\sim \left({\hbar^2\over G \Delta x}\right)^{1\over 3}
\ee

The qualitative behavior is as follows.
\bi
\item For $m\ll m_c$ is similar to a free particle; that is, the quantum spread dominates.
\item For $m \gg m_c$ the packet will undergo gravitational collapse.
\item For $m\sim m_c$ the wave packet is unstable.
\ei
In \cite{Carlip} it was estimated that the results are almost three orders of magnitude below any possible experimental reach. This presumably means that we will have experimental information in the not-too-distant future.

\section{Gravity as an emergent theory. Einstein's equations from thermodynamics.}
Recently \cite{Verlinde} (although some of the main  ideas are older than that \cite{Jacobson}) it has been emphasized that Einstein's equations can also be contemplated as emergent out of some more basic (unknown) degrees of freedom. From this point of view gravity would be like thermodynamics, or fluid mechanics; a macroscopic theory which makes no sense to extrapolate at arbitrary small lengths.
\par
Were these claims to be true, then the whole effort to quantize gravity is misplaced, and we should concentrate in the dynamics of those more basic degrees of freedom. Let us examine how this comes about with a little bit more detail. In string theory, gravity is also emergent, in a sense which in our opinion is not yet completely understood.
We will have more to say about this in the sequel.
\par
The starting point is Unruh's \cite{Unruh} fundamental observation that to every Killing horizon (which by definition is the locus of all points where the Killing becomes null, $\xi^2=0$) there is a temperature associated, which is proportional to the acceleration of the said Killing, which is defined through
\be
\xi^\l\nabla_\l \xi^\b=\kappa \xi^\b
\ee
This expression depends on the normalization of the Killing vector. Here we assume that it is fixed by demanding that $\xi^2=1$ asymptotically.
\par
This observation allows a thermodynamic interpretation of all horizons, somewhat similar to the one that Bekenstein and Hawking gave to the ordinary Schwarzschild's horizon.
\par
Let us then translate the thermodynamic formula
\be
\d Q= T d S
\ee
in the vicinity of a Killing horizon.
Consider a point $P$ at a horizon ${\cal H}$ with a corresponding Killing $\xi$.
\be
T={\hbar \kappa\over 2\pi}
\ee
where the acceleration of the normalized Killing can be computed through the convenient formula.
\be
\kappa^2 \equiv -{1\over 2}\nabla^\a\xi^\b\nabla_\a\xi_\b
\ee

 The flow of energy and momentum is given by the integral of the energy-momentum tensor
\be
\d Q=\int_{\cal H} T_{\a\b}\, \xi^\a\, d\s^\b
\ee
Call $k$ the tangent vector to the horizon and $\l$ the affine parameter (negative in the past of the point under consideration). Then
\be
\xi^\a=-\kappa \,\l\, k^\a
\ee
as well as
\be
d \s^\a=k^\a\, d\l\, dA
\ee
where $dA $ is the element of area on a cross section of the horizon.

\be
\d Q=-\kappa\int_{\cal H}\l\, T_{\a\b}\, k^\a\, k^\b \,d\l \, dA
\ee
Now the essential input is the assumption that the entropy associated to a Killing horizon is {\em not} extensive (that is, scaling with the volume); but that it rather scales with the area
\be
dS= \eta\d A
\ee
where $dA$ is {\em the area variation of a cross section of a pencil of generators of the horizon.}
It is a fact that
\be
\d A=\int_{\cal H}\theta\, d\l\,dA
\ee
where the expansion of the congruence, $\theta$, is definesd as
\be
\theta\equiv \nabla_\m k^\m
\ee
The dependence of this expansion is given by Raychaudhuri's equation 
\be
{d\theta\over d\l}=-{1\over 2}\theta^2-\s^2-R_{\a\b} k^\a k^\b
\ee
which can be trivially integrated, assuming that at the point chosen
\be
\theta=\s=0
\ee
yielding
\be
\theta=-\l\, R_{\a\b} k^\a\,k^\b
\ee
which in turn implies that
\be
\d A=-\int_{\cal H}\int \l\,R_{\a\b}\,k^\a\, k^\b\,d\l\,dA
\ee
Horizon thermodynamics then is telling us that
\be
T_{\a\b}\, k^\a\,k^\b={\hbar \eta\over 2\pi}\, R_{\a\b}\, k^\a\, k^\b
\ee
for all null vector $k$. This means that 
\be
T_{\a\b}={\hbar \eta\over 2\pi}\, R_{\a\b}+ f g_{\a\b}
\ee
Conservation of the energy momentum tensor then implies Einstein's equations
\be
R_{\a\b}-{1\over 2}\left(R+2\l\right)g_{\a\b}={2\pi\over \eta} T_{\a\b}
\ee
Consistency then demands that
\be
\eta={1\over 4\pi G}
\ee
In Erik Verlinde's version this argument is called {\em entropic}. The root of it, however, still lies in the hypothesis that the gravitational entropy is proportional to the area. 
The whole argument does not generalize easily to non-static situations, like for example, to the treatment of cosmology, in which time dependence is essential.
\par

In order to explain the existence of dark matter, Verlinde was forced to introduce a second set of quantum degrees of freedom, which are such that their entropy is extensive (that is proportional to the volume, like in normal laboratory systems).
\par
 This in turn allowed him to make some predictions on the static distribution of dark matter (no dynamics yet). It seems that for the time being observations are not supporting this alternative viewpoint. More tests are on the way though.
\par
At any rate, what these works prove is that independently of  what the fundamental quantum degrees of freedom could be,  black hole (or even horizon) thermodynamics leads to Einstein's equations in an almost unique way. Black hole physics acquires then an essential role as determining in some sense the whole dynamics of the gravitational field.
\section{Physics of Gravitons}
One of the smoking guns of a fundamental interaction is that there should be  an intermediate particle responsible for it; in our case, this would be the graviton. Physics of gravitons in flat space (and even in a curved background) can be studied by gauge theory techniques, as was pioneered long ago by Feynman and DeWitt, and culminating in the famous calculation to one loop by 't Hooft and Veltman \cite{thv}. 
This is still a very active topic of research. And this in spite of the fact that graviton cross sections are so small that there is no conceivable way in which they can be measured with present colliders. The reason is that this a way of making precise computations, and any inconsistency found here will presumably serve as a hint of what could happen in the non-perturbative regime as well as indications of the  region of parameter space in which this non-perturbative physics will be located. 
\par
There are indications that in some aspects gravity is similar to a double copy of a gauge theory. Amplitudes diverge much less than expected by naive power-counting arguments, and it is not excluded that the maximum supersymmetric theory with 32 supercharges (N=8 supergravity in 4 dimensions) could  be all loop finite.

We can parametrize our ignorance on the
{\em fundamental} ultraviolet (UV) physics by writing down all local operators of dimension $D$ in the low energy fields $\phi_i(x)$ compatible with the basic symmetries of all matter fields, represented schematically by $\phi_i$
\[
L=\sum_{D=0}^\infty{\l_D\over \Lambda^D}\,{\cal O}^{(D+4)}\left[\phi_i\right]
\]
Here $\Lambda$ is an ultraviolet (UV) cutoff, which damps the contributions of large euclidean momenta (or small euclidean distances) and $\l_D$ is an infinite set of dimensionless couplings.
\par
Standard Wilsonian arguments imply that {\em irrelevant operators}, corresponding to $D > 0$, which means that the total dimension of the operator is bigger than four,  are less and less important as we are interested in deeper and deeper infrared (IR) ({\em low energy}) variables.  The opposite occurs with {\em relevant operators}, corresponding to $D < 0$, so that the total dimension of the operator is less than four, like the masses, that become more and more important as we approach the IR. 
The intermediate role is played by the {\em marginal operators}, corresponding to precisely $D=0$, so that the dimension of the operator is exactly four, and whose relevance in the I R is not determined solely by dimensional analysis, but rather by quantum corrections. 
The range of validity of any finite number of terms in the expansion is roughly $E\leq \Lambda$.
where $E$ is a characteristic energy of the process under consideration.
\par
In the case of gravitation, we assume that general covariance (or diffeomorphism invariance) is the basic symmetry characterizing the interaction. In order to define fermions we need a locally inertial frame, $e_a^\m\pd_\m$ as well as a Lorentz (spin) connection, $\omega$ such that the spinorial covariant derivative is given by ${\cal D}\equiv \pd_\m +\omega_\m$. Then, somewhat symbolically

\bea
&L_{eff}=\sqrt{|g|}\bigg\{ \l_0 \Lambda^4 +\l_1 \Lambda^2 {\cal R} +\l_2 {\cal R}^2+ {1\over 2}\,g^{\a\b}\nabla_\a\phi\nabla_\b\phi+{\l_3 \over \Lambda^2}{\cal R}^{\a\b}\nabla_\a\phi\nabla_\b\phi+{\l_4 \over \Lambda^2}\,{\cal R}^3+\l_5 \phi^4\bigg\}+\nonumber\\
&+\bar{\psi}\,e^\m_a \g^a {\cal D}_\m\psi+{\l_5\over \Lambda^2}\bar{\psi}e^\m_a \g^a\, \cR\,{\cal D}_\m\psi+\ldots
\eea
We have represented by ${\cal R}$ any combination of the Riemann tensor and the spacetime metric, leaving implicit the detailed index structure .
If we aim  to recover General Relativity in the classical IR limit we are forced  to match 
\[
\l_1\Lambda^2=-{c^3\over 16\pi G}\equiv -2 M_p^2
\]
This in turn, means that if $\l_0\Lambda^4$ is to yield the observed value for the cosmological constant (which is of the order of magnitude of Hubble's constant, $H_0^4$, which is a very tiny figure when  expressed in particle  physics units, $H_0\sim 10^{-33}\,eV$) then
\[
\l_0\sim 10^{-244}
\]
This is one aspect of the cosmological constant problem; it seems most unnatural that the cosmological constant is observationally so small from the effective lagrangian point of view.
\par
 This fact can indeed be used as an argument against the whole effective lagrangians philosophy. I do not have anything new to say on this, except the obvious comment that before dismissing the whole idea one has to put on the other side of the balance the enormous successes of the effective lagrangian techniques in describing low energy physics both in QCD and in the electroweak model.
\par
This expansion is fine as long as it is considered a low energy expansion. As Bjerrum-Bohr, Donoghue and Holstein \cite{BjerrumBohr} \cite{Donoghue} have emphasized, even if it is true that each time that a renormalization is made there is a finite arbitrariness, there are physical predictions stemming from the non-local finite parts.
\par
At any rate,  when energies are reached that are comparable to Planck's mass,
\[
E\sim M_p.
\]
Then all couplings  in the effective Lagrangian become of order unity, and there is no {\em decoupling limit} in which gravitation can be considered by itself in isolation from all other interactions.
This then seems the minimum prize one has to pay for being interested in quantum gravity; all couplings in the derivative expansion become important simultaneously.
\par
 No significant differences in this respect appear when supergravity is considered.
 \par
  The root of the problem lies in the assumption of Diff invariance, which affects all spacetime fields. This assumption in turn has its roots on the equivalence principle, which implies the existence of LIF (locally inertial frames) at every point of the spacetime manifold.
\par
On the other hand, it used to be thought that all QFT involving gravity were necessarily divergent in a non-renormalizable way, even when considering gravitons in flat space. For pure Einstein-Hilbert gravity this was first shown by Goroff and Sagnotti \cite{Goroff} who found a divergent piece at two loops that did not vanish on-shell. For example, it was believed that N=8 supergravity (a theory with 32 supercharges) would be divergent starting at 3 loops ($10^{20}$ diagrams approximately would have to be computed to check this).
\par

Recent advances in the computation of QFT amplitudes however, (mainly by Zvi Bern and coworkers \cite{Bern}) have unveiled unsuspected cancellations, that can not be explained by known symmetries.
\par
 The key idea of this research was  to built on shell amplitudes starting solely from three-particle vertices, making thus irrelevant all the complicated higher point vertices that used to cloud quantum gravity computations. This idea in turn was inspired by string theory, where gravitons appear in the closed string sector, and gauge fields in the open string sector. 
 \par
Another important insight (stemming also from string theory) was that in many respects, gravity amplitudes behave as a double copy of the much simpler gauge theory amplitudes.
\par
At tree level  there is a duality of sorts between color factors on the gauge side and kinematic factors in quantum gravity, Symbolically, if a gauge amplitude is
\be
A=i g^{n-2}\sum_i {c_1 n_i\over D_i}
\ee
where $c_i$ are color factors, $n_i$ are kinematics factors, and $D_i$ propagator denominators, then the corresponding gravity amplitude is
\be
M=i \kappa^{n-2}\sum{n_i^2\over D_i}
\ee
Now there seems to be a consensus that the first divergences will appear in N=8 supergravity (Sugra) not before 7 loops. Some people even believe that the theory could be all loop finite.
\par
Bern's group \cite{Bern} has recently proved the finitness of this very theory up to 5 loops; the critical dimension at which the first divergence appears is $D_c={24\over 5}$.
\par
At any rate, there is ample evidence in gauge theories (QCD in particular) that the non-pertubative sector is quite important (in the case of QCD it dominates the low energy infrared regime). 
\par
It is likely that non-perturbative effects will be even more complicated in quantum gravity, to the extent that  string theory dualities \cite{Hull} serve as some indication of the true physics.
\par
 Nevertheless, all the results  by Zvi Bern and collaborators are on the ultraviolet behaviour of N=8 supergravity can be  explained by supersymmetry cancellation when appropriately applied \footnote{
 I am grateful to Pierre Vanhove for pointing out these facts to me.} \cite{GRV}.

The absence of 3-loop divergence was a prediction of the string theory dualities before the QFT computation was done.  
The fact that supersymmetry allows a 5-loop divergence in D=24/5 was predicted in \cite{GreeBj} \cite{Vanhove2}.
In these works the coefficients had not been computed so it could have happened that the divergence is absent after all. The absence of divergence at 5-loop would have been there an indication that the UV property of N=8 supergravity cannot be explained by supersymmetry or by any known symmetries of quantum gravity.  But the divergence is present so this means that no extra symmetry is at  work in this case at least.
\par
Besides,  the counter-term of the ultraviolet divergence up-to and including 3-loop order has been derived from string theory.  The counter-term is part of the U-duality invariant coupling of the low-energy expansion of string theory \cite{Pioline}.

\subsection{The background field approach in quantum field theory.}
Bryce DeWitt \cite{DeWitt} pioneered the use of covariant methods in QFT. Thanks to those methods we can, for example study quantum fluctuations around an {\em arbitrary} gravitational background. The technique he introduced is precisely named  the {\em background field method} which later on found applications in some gauge theory computations as well \cite{Abbott}. We can give here but a superficial glimpse of what it is about. Consider the vacuum persistence amplitude in the presence of an arbitrary source (sometimes called the {\em  partition function} by  analogy with statistical mechanics),
\be
Z[J]\equiv e^{i W[J]}\equiv \int {\cal D} \phi~ e^{i S[\phi]+ i \int d(vol)\,J(x)\phi(x)}
\ee
Where in this formal analysis we represent all fields (including the gravitational field) by $\phi(x)$, and we add a coupling to an arbitrary external source as a technical device to compute Green functions out of it by taking functional derivatives of $Z[J]$ and then putting the sources equal to zero. This trick was also invented by Schwinger (DeWitt's former advisor). The partition function generates all Green functions, connected and disconnected. Its logarithm, $W[J]$ sometimes dubbed the {\em free energy} (this nam ealso  comes from a direct analogy with similar quantities in statistical physics), generates connected functions only.
\par
It is possible to give an intuitive meaning to the path integral in quantum mechanics as a transition amplitude from an initial state to a final state. This is actually the way Feynman derived it. 
\par
In quantum field theory (QFT) the integration measure is not mathematically well-defined.
For loop calculations, however, it is enough to {\em formally  define} the gaussian path integral as a functional determinant, that is
\be
\int {\cal D}\phi\, e^{i \langle\phi |K \phi\rangle}=\left(\text{det}~K \right)^{-{1\over 2}}
\ee
where the scalar product is defined as 
\be
\langle\phi| K \phi\rangle\equiv \int d(vol)~\phi~ K~\phi
\ee
and $K$ is a differential operator, usually
\be
K=\nabla^2+\text{something}
\ee
There are implicit indices in the operator to pair the (also implicit) components of the field $\phi$.
\par
The only extra postulate needed is translation invariance of the measure, in the sense that
\be
\int {\cal D}\phi\, e^{i ~\left\langle \left(\phi+\chi\right)~\left| K~ \right.\left(\phi+\chi\right)\right\rangle)}=\int {\cal D}\phi\, e^{i \left\langle\phi~| K ~\phi\right\rangle}
\ee
This is the crucial property that allows the computation of integrals in the presence of external sources by completing the square.
\par
It is quite useful to introduce a generating function for one-particle irreducible (1-PI)  Green functions. This is usually called the {\em effective action} and is obtained through a Legendre transform, quite analogous to the one  performed when passing from the Lagrangian to the hamiltonian ion classical mechanics. 
\par
One defines the {\em classical field} as a functional of the external current by
\be
\phi_c[J]\equiv {1\over i}~{\d W[J]\over \d J(x)}
\ee
This equation allows, by the implicit function theorem, the formal definition of the inverse function, $J=J[\phi_c]$. 
The Legendre transform then reads
\be
\Gamma[\phi_c]\equiv W[J]-i \int d^n x  J(x)\phi_c(x)
\ee
It is a fact that
\be
{\d \Gamma\over \d \phi_c(x)}=\int d^n z ~{\d W\over \d J(z)}~{\d J(z)\over \d \phi_c(x)}-i J(x)-i \int d^n z \phi_c (z)~{\d J(z)\over \d \phi_c(x)}=- i J(x)
\ee

Let us introduce now the background field technique first in the language of Yang-Mills theories.
The main idea is to  split the integration fields into a {\em classical} and a {\em quantum} piece:
\be
W_\m\equiv \bar{A}_\m+ A_\m
\ee
The functional integral is performed over the quantum fields only.
where for an ordinary gauge theory the action has three pieces. First the gauge invariant piece
\be
L_{\text{gauge}}\equiv -{1\over 4} F_{\m\n}[W]^2
\ee
with
\be
F^a_{\m\n}[W]\equiv \pd_\m W^a_\n-\pd_\n W^a_\m+ g f_{abc}W^b_\m W^c_\n 
\ee
where as usual,
\be
\left[T_a,T_b\right]=i f_{abc} T_c
\ee
The gauge transformations with parameter $\omega^a$ are
\be
\d W_\m^a\equiv \d\left(\bar{A}^a_\m+ A^a_\m\right)\equiv -f_{abc}\, \omega^b\, \,W^c_\m +{1\over g} \pd_\m \omega^a=-f_{abc}\, \omega^b \left(\bar{A}^a_\m+ A^a_\m\right) +{1\over g} \pd_\m\omega^a
\ee
This can be implemented in two ways. First letting the background field be inert. Those are the {\em quantum gauge} 
\bea
&&\d_Q \bar{A}_\m^a\equiv 0\nonumber\\
&&\d_Q A_\m^a=-f_{abc} \omega^b \left(\bar{A}_\m^a+ A^a_\m\right) +{1\over g} \pd_\m\omega^a
\eea
Those are the transformations in need of gauge fixing. It is to be remarked that gauge symmetry is realized non-linearly on the quantum fields. 
\par
It is also possible to reproduce the total gauge transformations through the {\em classical} transformations
\bea
&&\d_C \bar{A}_\m^a=-f_{abc} \omega^b \bar{A}^a_\m +{1\over g} \pd_\m\omega^a\nonumber\\
&&\d_C A^a_\m=-f_{abc} \omega^b A^c_\m
\eea
under which the quantum field transforms linearly as an adjoint  vector field.
\par
Currents transform in such a way that
\be
\d_C \int J_a^\m A^a_\m=0
\ee
that is
\be
\d J^a_\m=- f_{abc} \omega^b J^c_\m
\ee
The beauty of the background field method is that it is possible to gauge fix the quantum symmetry while preserving the classical gauge symmetry. All computations are then invariant under gauge transformations of the classical field, and so are the counterterms. This simplifies the heavy work involved in computing Feynman diagrams in the presence of dynamical  gravity.
\par

The simplest background gauge is
\be
\bar{F}^a[A]\equiv \pd_\m A_a^\m + g f_{abc} \bar{A}^b_\m A^\m_c\equiv \left(\bar{D}_\m A^\m \right)^a
\ee
L. Abbott \cite{Abbott} was able to prove a beautiful theorem to the effect that the effective action computed by the background field method is simply related to the ordinary effective action
\be
\Gamma_{BF}[A^{BF}_c,\bar{A}]=\left.\Gamma[A_c]\right|_{A_c=A_c^{BF}+\bar{A}}
\ee
This means in particular, that
\be
\Gamma[A_c]=\Gamma_{BF}[0,\bar{A}=A_c]
\ee
At the one loop order all this simplifies enormously. Let us spell in detail the simplest scalar case. Working in euclidean space, introducing sources for the quantum fields only, and in a schematic notation,
\bea
&e^{-W[\bp]}\equiv \int {\cal D}\phi\, e^{-S[\bp]-\int d(vol)\, \phi\, K[\bp]\,\phi-\int d(vol)\, J\,\phi}=\nonumber\\
&=e^{-S[\bp]-{1\over 2}\text{log~det}~K[\bp]-{1\over 4} \int d(vol)\, J\, K^{-1}[\bp]\, J}
\eea
This means that the classical field in the background field formalism is given by
\be
\phi_c=-{1\over 2} K^{-1}~[\bp] J
\ee
so that the sources read
\be
J= -2 K[\bp]~\phi_c
\ee
and the action can be written as
\bea
&\Gamma^{BF}[\phi_c,\bp]=W[J(\phi_c)]-\int d(vol)\,J\phi_c=\nonumber\\
&=S[\bp]+{1\over 2}\text{log~det}~K[\bp]+ \int d(vol)\, \phi_c\,K[\bp]\,  K^{-1}[\bp] \,K[\bp]\, \phi_c-2 \int d(vol)\,\phi_c\,   K[\bp]\, \phi_c=\nonumber\\
& =S[\bp]+{1\over 2}\text{log~det}~K[\bp]-\int d(vol)\,\phi_c\, K[\bp]\, \phi_c
\eea
Then by Abbott's theorem
\be
\Gamma(\phi_c)=\Gamma^{BF}[0,\bp=\phi_c]=W[\bp]\equiv S[\bp]+{1\over 2}\text{log~det}~K[\bp]
\ee
The one loop effective action is equal to the background field free energy, and the background field can be equated to the classical field.
\par
The case of gravitations is analogous (but for algebraic complexity), We again split the gravitational field as
\be
g_{\m\n}=\bg_{\m\n}+\kappa \,h_{\m\n}
\ee
This is done so that the mass dimension of the graviton field $h_{\m\n}$ is equal to one. The full diffeomorphism  gauge symmetry
\bea
&\d g_{\m\n}\equiv \d \left(\bg_{\m\n}+\kappa h_{\m\n}\right)=\nabla_\m\xi_\n+\nabla_\n\xi_\m=\nonumber\\
&=\pounds(\xi) g_{\m\n}\equiv \xi^\l\pd_\l g_{\m\n}+g_{\m\l}\pd_\n\xi^\l+g_{\l\n}\pd_\m\xi^\l
\eea
includes both the one-loop background field transformations
\bea
&\d \bg_{\m\n}=\pounds(\xi)\bg_{\m\n}\nonumber\\
&\d h_{\m\n}={1\over \kappa}\left(\bn_\m\xi_\n+\bn_\n\xi_\m\right)
\eea
(the term $\xi^\l\pd_\l h_{\m\n}$ is of two-loop order). It also includes the quantum gauge transformations
\bea
&\d\bg_{\m\n}=0\nonumber\\
&\d h_{\m\n}=\bn_\m\xi_\n+\bn_\n\xi_\m
\eea
A convenient gauge fixing is the {\em harmonic} or {\em de Donder} gauge, defined by
\be
\bn_\m h^{\m\n}={1\over 2}\,\bn^\m h
\ee
(where $h\equiv \bn^{\a\b} h_{\a\b}$). The reason for the name is as follows. Consider the coordinates as functions.
Then
\be
\Box x^\m={1\over \sqrt{-g}}\pd_\a\left(\sqrt{-g}\,g^{\a\m}\right)=-\Gamma^\m_{\r\s}\,g^{\r\s}
\ee
When perturbing around flat space
\be
g_{\m\n}\equiv \eta_{\m\n}+\kappa h_{\m\n}
\ee
it so happens that
\be
\Gamma^\m_{\r\s}\,g^{\r\s}={1\over 2}\left(-\pd^\m h+ 2 \pd_\l h^{\m\l}\right)
\ee
and the gauge condition is equivalent to the demand that the coordinates are harmonic when considered as functions on the manifold.
\par
In conclusion, there is a systematic way of studying quantum fluctuations around an arbitrary background, at least as long as this background is a solution of the classical Einstein's equations. This should be enough to study quantum corrections to gravitational waves, for example.
\par
It is not yet excluded that some supersymmetric versions of this problem can lead to finite theories, at least when the background is flat.
\par
Some non-generic subtle points appear in the frequent case when the background has got horizons and/or singularities.

\section{What are the questions whose response we are seeking for?}
Let us think for a moment what are the questions we would like to be answered by our theory of quantum gravity.
\par
We are used in quantum mechanics or quantum field theory (QFT) to be able to answer questions on probability amplitudes for transitions between states at given times (or given spacetime points).

Related computations involve correlators of strings of spacetime fields acting on the vacuum. Reduction techniques relate those correlators to S matrix elements, which are directly related to cross sections that can be observed in colliders.
\par
In some Hilbert space of sorts (that is, a Fock space) , Sch\"odinger's equation  (or equivalently, Heisenberg's equation of motion) still holds. Those relate the time derivative of either the physical state or else the spacetime fields to some second member involving the full hamiltonian of the system.

It is quite difficult to generalize any of these concepts to quantum gravity. This is specially so if we insist (as we do) in the correspondence principle alluded to formerly; that is, demanding a smooth limit when $\kappa\rightarrow 0^+$
\bi
\item 
The first thing is that there is no a natural definition of time in GR (and correspondingly of a hamiltonian). We shall review in a moment why this is so.

\item Besides, the starting point for the analysis of causality  (and unitarity, more on this later) in QFT is the {\em microcausality condition}, namely
\be
\left(x-y\right)^2 < 0 \Longrightarrow \left[\phi(x),\phi(y)\right]=0
\ee
that is, fields at spacelike separated points do commute. It is not know what is the correct generalization of this idea to GR, or even if there is a natural notion of causality there.

\item As a consequence, there is no clear definition of causality. At the classical level, it can be related to conjectures that aim at keeping a well-defined Cauchy problem, such as the {\em Hawking's chronology conjecture} (no closed timelike curves) or  {\em Penrose's cosmic censorship} (no naked singularities), but even those ideas are not easy to generalize at the quantum level.
\item
 Something similar happens with the related concept of unitarity. In QFT it is believed to be equivalent to
 \be
 S S^+=1
 \ee
 but in general spacetimes (i.e., not asymptotically flat) there is no concept of S-matrix, so that it is not clear what is the physical meaning of amplitudes or even S-matrix.

\item Finally, when quantum gravity is applied to the whole universe (quantum cosmology) besides the conceptual question of  what is the observable meaning of a transition amplitude between two different 3-space geometries, (we do not have a set of universes at our disposal) there is unavoidably the general question of how to generalize the Copenhagen interpretation of quantum mechanics \cite{Hartle}, because there is no possible distinction between the observer and the system. Then, we are in a real swampland. 
\par
What is the physical meaning of the assertion that the probability of transition from one 3-space geometry $\Sigma_1$ to another one $\Sigma_2$ at some other cosmic time $T$ is a given number $P(\Sigma_2,\Sigma_1,T)$? How can we verify those  hypothetical predictions, even in principle?

It is quite often mentioned that quantum gravity would give information on space-time singularities, both inside black holes (where they are hidden behind a horizon) and at the big bang. which is classically a naked singularity.
\par

It may well be so. But not necessarily.
\par
 It could be the case that quantum gravity yields no information beyond the platitude that in quantum mechanics geometrical points are in contradiction with the uncertainty principle.
\par
 For example, in quantum electrodynamics (QED) Coulomb's singularity is hidden by the renormalization procedure, a procedure that tames all divergences, but one that gives no particularly physical intuition on them, other that the dependence of the fine structure constant  with the renormalization scale, as determined by the renormalization group.
\ei
 \section{ The canonical approach.}

It is widely acknowledged that there is a certain tension between a $(3+1)$ decomposition
implicit in any canonical approach, privileging a particular notion of time, and the
 beautiful geometrical structure of general relativity, with its invariance under general
coordinate transformations. 
\par
To begin with, it is clear that any hypothetical definition of energy cannot have any tensorial character. In fact, the equivalence principle guarantees that there is a LIF (locally inertial frame) in which there is no gravitational field (again, locally) and consequently, no energy associented to it.
\par
Let us now nevertheless explore how far we can go on this road.
\subsection{Quasilocal energy, pseudotensors and superpotentials.}
Before beginning with the canonical formalism proper, let us make a few general remarks on the concept of energy in general relativity (GR). The definition of energy in GR was (and still is) one of the main problems that kept physicists and mathematicians alike busy for many years, including Einsteim, Hilbert, Noether, Klein, Dirac, and so on.
And this in spite of tha fact that it is clear that the equivalence principle tells us immediatly the there cannot be any tensorial quantity with that physical interpretation. The reason is that at any point of spacetime one can choose a Local Inertial Frame (LIF), in which the gravitational field vanishes (locally) and so must do its energy.
\par
It is interesting to remind us of some of those efforts. They have been nicely reviewed in \cite{Chang}.

Let us start by choosing a particular frame  as well as  a superpotential (nothing to do with supersymmetry)
\be
H_\m^{\n\l}=H_\m^{[\n\l]}
\ee
and split Einstein's tensor in such a way that
\be
\kappa \sqrt{|g|} N^\m t^\l_\m\equiv -N^\m \sqrt{|g|} G_\m^\l+{1\over 2} \pd_\s\left(N^\m H_\m^{\l\s}\right)
\ee
Now we particularize to a frame in which the components of the vector $N^\m $ are constant
\be
\pd_\l N^\m=0
\ee
Then, in this frame, and using Einstein's equations
\be
G_{\m\n}=\kappa\, T_{\m\n}
\ee
we get
\be
\pd_\l H_\m^{\n\l}=2\kappa \sqrt{|g|}{\, \cal T}_\m^\n\equiv2\kappa\sqrt{|g|}\left(t_\m^\n+T_\m^\n\right)
\ee
Owing the the skewness of the superpotential, the total energy-momentum is conserved in the ordinary, non-covariant sense
\be
\pd_\l\left(\sqrt{|g|}\, {\cal T}_\m^\l\right)=0
\ee
This then yields a conserved energy-momentum
\be
N^\m {\cal P}_\m\equiv \int N^\m\, {\cal T}_\m^\n \, d\Sigma_\n
\ee
In particular, the energy in a spacelike volume  $ V $ is given by
\bea
&P_N (V)\equiv \int_V N^\m\, {\cal T}_\m^\n \, d\Sigma_\n={1\over 2\kappa}\int_V\pd_\l\left(N^\m H_\m^{\n\l}\right)\,d\Sigma_\n=\nonumber\\
&=\int_{\Sigma=\pd V}\,{\cal B}^{\n\l}(N)\,d\sigma_{\n\l}
\eea
where
\be
{\cal B}^{\n\l}(N)\equiv {1\over 2\kappa}\,N^\m H_\m^{\n\l}
\ee
This has led to the authors of \cite{Chang} to conclude that for any pseudotensor, the associated superpotential is naturally a hamltonian boundary term.
This energy so defined is {\em quasilocal} in the sense that it does not depend on the values of the frame and the fields on the whole volume whose energy is being computed; but only on the corresponding values at the boundary of said volume. There is no simple way in which this is conserved, though, except in the familiar Arnowitt-Deser-Misner (ADM) case \cite{ADM} of asymptotically flat spacetimes.
\par
Let us examine a concrete example that will lead to M\o ller's pseudotensor.
Define the connection 1-forms
\be
\omega^\a\,_\b\equiv \Gamma^\a_{\b\g} dx^\g
\ee
and the corresponding curvature 2-form
\be
\Omega^\a\,_\b\equiv d\omega^\a\,_\b+\omega^\a\,_\g\wedge\omega^\g\,_\b
\ee
Then the Einstein-Hilbert lagrangian can be written in the somewhat pedantic form
\be
L_{EH}=\Omega^\a\,_\b\wedge\eta_\a\,^\b \equiv \Omega^\a\,_\b\wedge*\left(dx_\a\wedge dx^\b\right)=R\,d(vol)
\ee

The Hamiltonian can be defined as usual
\be
i_N L=i_N \Omega\wedge \eta+\Omega\wedge i_N \eta=\pounds_N \omega\wedge \eta-{\cal H}(N)
\ee
where
\bea
&-{\cal H}(N)=-d i_N\omega\wedge \eta+i_N\left(\omega\wedge \omega\right)\wedge \eta+\Omega\wedge i_N\eta=\nonumber\\
&=-d \left(i_N\omega\wedge \eta\right)+i_N\omega\wedge d\eta+i_N\left(\omega\wedge\omega\right)\wedge \eta+\Omega\wedge i_N\eta
\eea
First of all, define
\be
\eta\equiv \e_{a_1\ldots a_n}e^{a_1}\wedge\ldots e^{a_n}
\ee
Then, for the metric connection
\be
D\eta\equiv d\eta+\sum \e_{a_1\ldots a_n}\,\omega^{a_1}\,_c\, e^{c}\wedge\ldots e^{a_n}=0
\ee
owing to the condition
\be
d e^a+\omega^a\,_b\wedge e^c=0
\ee
Let us now examine
\be
i_N\omega\wedge d\eta+i_N\left(\omega\wedge\omega\right)\wedge \eta=i_N\omega\wedge D\eta=0
\ee
It is useful to distinguish between  g
\be
N^\m {\cal H}_\m\equiv -\Omega^\a\,_\b\wedge N^\m \eta_\a\,^\b\,_\m
\ee
which is claimed to project to the usual ADM hamiltonian, and the boundary term
\be 
B(N)\equiv i_N \omega\wedge \eta=N^\l \omega^\a\,_{\b\l} \eta^\b\,_\a\equiv N^\l {\cal M}^{\a\b}\,_\l\, dS_{\a\b}
\ee
with
\be
dS_{\a\b}\equiv {1\over 2}\,\e_{\a\b\g\d}\, dx^\g\wedge dx^\d
\ee
To be specific
\be
{\cal M}^{\a\b}\,_\l\,=\sqrt{|g|}\,\left(g^{\s\b}\Gamma^\a_{\s\l}-g^{\s\a}\Gamma^\b_{\s\l}\right)
\ee

This is the {\em superpotential} (again, nothing to do with supersymmetry) whose divergence yields {\em M\o ller's pseudotensor}.

It is not clear what is the conclusion of this canonical analysis. It was clear since the beginning that there is a deep contradiction between  the hamiltonian  being a geometrical quantity on the one hand, and the principle of general covariance (or diffeomorphism invariance)
which forbids privileging certain frames of reference.
On the other hand, all known formulations of quantum mechanics are related to the hamiltonian in an essential way. 
There seems to be no easy way out. On the one hand, to insist that some frames of reference are special in some sense (like Fock's harmonic coordinate systems) is ugly and goes against the beautiful GR philosophy. On the other hand, as we have already seen, it is very difficult to modify, whatever slightly, quantum mechanics.

The modern theory of hamiltonians for systems with constraints is rooted in the analysis of General Relativity. Anderson and Bergmann \cite{Anderson} introduced the concepts of primary and secondary constraints, as well of first class and second class constraints. Al this was beatifully explained by Dirac \cite{Dirac} in his famous Yeshiva lectures.
Thereby it was noticed that for a system such that the lagrangian is homogeneous of first order in the velocities, that is
\be
\sum_{i=1}^N \dot{q}^i{\pd L\over \pd \dot{q}^i}= L
\ee
the naive hamiltonian vanishes, because
\be
H\equiv \sum_i p_i \dot{q}^i-L=0
\ee
The first thing to notice is that if we redefine the time coordinante as
\be
t\rightarrow \t(t)
\ee
then the action is invariant,
\be
\int dt\, L\left(q,{d q\over dt}\right)=\int d\t\, L\left(q,{d q\over d\t}\right)
\ee
This means that the time variable can be chosen at will, the action is insensitive to this choice.
\par
The second thing to notice is that necessarily the $p_i$ are homogeneous functions of degree zero of the velocities. That is, they are  functions of the ratios of velocities, of which there are only $N-1$ independent ones. This  means that there is at least one primary constraint.
The total hamiltonian consists in the sum of all primary first class constraints multiplied with arbitrary coefficients
\be
H_T\equiv \sum_a v_a \phi_a
\ee
The EM would read
\be
\dot{g}\sim\sum_a v_a \left\{g,\phi_a\right\}
\ee
It is clear that, in spite of the fact that we are in the framework of Newtonian mechanics,  there is no absolute time variable here. Any monotonic function of $t$ is valid as a new time variable. There is no absolute time; this is in agreement with our previous observations.
\par
Now it is a fact that {\em any} theory can be put into this form just by defining the time variable as an extra coordinate $t\equiv q^0$, and introducing a new time variable, $\t$. 
\be
L_{new}\left(q^0,q^i,{d q^0\over d\t},{d q^i\over d\t}\right)\equiv L\left(q^i,{{d q^i\over d\t}\over {d q^0\over d\t}}\right)\,{d q^0 \over d\t}
\ee
It is plain that the action is invariant
\be
S\equiv \int L_{new}\, d\t=\int L \, dt
\ee

\subsection{Dirac universal brackets}

In this section it will prove convenient to reserve the label $y^\a$ for the spacetime coordinates in Minkowski space to tell them apart from  the $3+1$ coordinates $(t,x^i)$ to be defined in the sequel.

Let us  now  consider a foliation of space-time given by the function
\be
t(y^\a)=C
\ee
The level hypersurfaces are spacelike; that is, the normal vector
\be
n_\a\equiv  N\pd_\a t
\ee
 is timelike, and will always be assumed normalized
 \be
 n^2=1
 \ee
 so that
 \be
{1\over  N^2} \equiv g^{\a\b}\,\pd_\a t\, \pd_\b t
 \ee
We shall also assume the existence of a $\infty^3$ congruence of curves
\be
y^\a\equiv \s^\a(x^i,t)
\ee
Each curve in the congruence is thepresented by
\be
x^i= C^i
\ee

 Tangent vectors on the $t=constant$ hypersurface are given by
\be
\xi_i^\a(t)\equiv \pd_i\s^\a(t)\quad i=1\ldots 3
\ee
The normal vector $n$ is orthogonal to all tangent vectors
\be
g_{\a\b}\,\xi_i^\a\, n^\b=0\quad i=1,\ldots 3.
\ee
The  vector tangent to the congruence is not necessarily normal to the hypersurfaces; it can be expanded in general as 
\be
N^\a\equiv{\pd \s^\a\over \pd t}\equiv N n^\a+ N^i \xi_i^\a\equiv N n^\a+ {\cal N}^\a
\ee
where 
\be
N\equiv N^\a n_\a=~{\pd \s^\a\over \pd t}~N~\pd_\a t
\ee
 is usually denoted as the {\em lapse} and $N^i$ is the {\em shift} in ADM's (Arnowitt-Deser-Misner) notation.
 \par
  Again, this means that the vector that goes from the point $(t,x)\in\Sigma_t$ to the point $(t+dt,x)\in \Sigma_{t+dt}$ does not lie necessarily in the direction of the normal to the hypersurface $t=constant$.
\par
Actually from consistency of the previous  definitions it  follows that
\be
N^\a\pd_\a t=1
\ee
Also our parametrization of the curves of the congruence as $x^i=C^i$ imply that
\be
\pounds(N^\a)\xi^\b_i=0
\ee
Finally, the fact that the coordinates $x^i$ on each surface are independent means that, considered as spacetime vectors, $\vec{\xi}_i\equiv \xi_i^\a\pd_\a$
\be
\left[\vec{\xi}_i,\vec{\xi}_j\right]=0
\ee

The $\infty^3$ dymamical variables $\s^\a$ will have some canonically conjugate momenta
\be
\{\s^\m(t,x),\pi_\n(t,y)\}=\d^\m_\n \d^{3}(x-y)
\ee
If the generalized coordinates $\s^\m$ are to vary at all in the dynamics, the constraints have to involve the conjugate momenta, so that it must be possible to write the constraints as
\be
H_T=\int d^{3}x\quad c^\a(t,x) \left(\pi_\a+K_\a\right)
\ee
Then it is a fact that
\be
\dot{\s}^\m\equiv \{\s^\m,H_T\}=c^\m
\ee

\vspace{1cm}

It is useful to decompose any vector index into normal and tangential components
\bea
& V_n\equiv V.n\equiv g_{\a\b} V^\a n_\b \nonumber\\
& V_i\equiv V.\xi_i\equiv g_{\a\b} V^\a \xi_i^\b
\eea
There is always the danger of taking $V_i$ such defined as the space components of the n-dimensional quantity $V$, but we shall try not to do so in the sequel. 
Actually,
\be
V^\m= V_n.n^\m+V_i h^{ij}\xi_j^\m
\ee
provided we define as usual the {\em induced metric}  on the hypersurface as 
\be
h_{ij}\equiv \xi^\m_i\, g_{\m\n}\,\xi^\n_j
\ee
which is nonsingular if the hypersurfaces are everywhere spacelike. Then we define the inverse matrix 
\be
h^{ij}h_{jk}\equiv\d^i_k
\ee
(Please notice that  in general, when the shift does not vanish,
\be
h^{ij}\neq g^{ik} g^{lm} h_{lm})
\ee
The spacetime metric can be reconstructed out of the lapse and shift through
\bea
&&ds^2=g_{\m\n} dy^\m dy^\n=g_{\m\n}\left(N^\m dt+ \xi^\m_i dx^i\right)\left(N^\n dt+ \xi^\n_i dx^i\right)=\nonumber\\
&&=N^2 dt^2+h_{ij}\left(dx^i+ N^i dt\right)\left(dx^j+N^j dt\right)
\eea
This in turn implies
\be
g_{\m\n}=n_\m n_\n+\xi_\m^i\, \xi_{\n i}
\ee

Then Dirac showed that for all these systems there is an universal set of Poisson brackets, namely,
\bea
&\{\pi_r(t,x),\pi_s (t,x^\prime)\}=\pi_s(t,x)\pd_r\d(x-x^\prime)+\pi_r (t,x^\prime)\pd_s \d(x-x^\prime)\nonumber\\
&\{\pi_n(t,x),\pi_r(t,x^\prime)\}=\pi_n(t;x^\prime)\pd_r \d(x-x^\prime)\nonumber\\
&\{\pi_n(t,x),\pi_n(t,x^\prime)\}=-2 \pi^r(t,x) \pd_r(x-x^\prime)-\Delta \pi(t,x) \d(x-x^\prime) d
\eea
where we define to save space
\be
\d(x-\xp)\equiv \d^{(3)}(x-\xp)
\ee
and as usual the ordinary thee-dimensional laplacian is
\be
\Delta\equiv \d^{ij}\pd_i\pd_j
\ee

Let us work this out in some detail.
\bea
&0=\{n_\m \xi^\m_i, \pi_{\n^\prime}\}=\{n_\m,\pi_{\n^\prime}\}\xi^\m_i+n_\m\{\xi^\m_i,\pi_{\n^\prime}\}=\nonumber\\
&=\{n_\m,\pi_{\n^\prime}\}\xi^\m_i+n_\n \pd_i\d(x- x^\prime)
\eea
We learn that
\be
\{n_\m,\pi_{\n^\prime}\}\xi^\m_i=-n_\n\pd_i\d(x -x^\prime)
\ee
Again,
\be
0={1\over 2}\{n^2,\pi_{\m^\prime}\}=\{n_\m,\pi_{\m^\prime}\}n^\m
\ee
Then
\bea
&\{n_\l,\pi_{\nu^\prime}\}=\{n_\r,\pi_{\nu^\prime}\}\left(n^\r n_\l+\xi^\r_j\xi_{\l k} h^{jk}\right)=\nonumber\\
&=-n_\n\, \xi_\l^j\pd_j \d(x-\xp)\equiv -n_\n \d_{-\l}(x-\xp)
\eea
Notice that we have defined
\be
\d_{-\l}(x-\xp)\equiv\xi_\l^j\pd_j \d(x-\xp)
\ee
It follows that
\bea
&\{n_\l,\pi_{n^\prime}\}=\{n_\l,\pi_{\m^\prime}n^{\m^\prime}\}=-n^{\m^\prime} n_\m \d_{-\l}=\nonumber\\
&=-\xi^i_\l\pd_i\left(n^{\m^\prime}n_\m \d(x-x^\prime)\right)+\xi^i_\l \pd_i\left(n^{\m^\prime}n_\m \right)\d=-\xi^i_\l \pd_i \d(x-x^\prime)\nonumber
\eea
Before going on, let us show an elementary relationship. It is plain that
\be
\pd_j(n_\m \xi_i^\m)=0=\pd_j n_\m \xi^\m_i+ n_\m \pd_j \xi^\m_i
\ee
as well as
\be
n_\m \pd_i n^\m=0
\ee
also
\be
\pd_i n_\m\xi^\m_j=-n_\m \pd_i \xi^\m_j=-n_\m \pd_j \xi^\m_i=\pd_j n_\m \xi^\m_i
\ee
and multiplying by $\xi^{j\r}$
\be
\pd_i n^\r=\xi^{j\r}\pd_j n_\m \xi^\m_i
\ee
Let us now define the construct $n_{\m-\n}$
\be
n_{\m-\n}\equiv \pd_i n_\m \xi^i_\n=\xi^i_\n \xi^{j\m}\pd_j n_\l \xi^\l_i=\xi^{j\m}\pd_j n_\n\equiv n_{\n-\m}
\ee
It follows that
\bea
&\{n_\l,\pi_{j^\prime}\}=\{n_\l,\pi_{\nu^\prime}\xi_{j^\prime}^{\n^\prime}\}=-n_\n \xi_\l^k\pd_k \d(x-x^\prime)\xi_{j^\prime}^{\n^\prime}=\nonumber\\
&=-\pd_k\left(n_\n \xi^{\n^\prime j^\prime} \xi_\l^k \d(x-x^\prime) \right)+\pd_k\left(n_\n \xi_\l^k \xi_{j^\prime}^{\n^\prime}\right)\d(x-x^\prime)=
\pd_k n_\n\xi_\l^k \xi_{j}^{\n}\d(x-x^\prime)=\nonumber\\
&=\pd_k n_\l \xi_\n^k \xi_j^\n(x-x^\prime)=\pd_j n_\l \d(x-x^\prime)\nonumber
\eea
 Finally
 \bea
& \{\pi_n,\pi_{n^\prime}\}= \{\pi_\l n^\l,\pi_{\m^\prime} n^{\m^\prime}\}= \pi_\l n^{\m^\prime}\{n^\l,\pi_{\m^\prime}\}+ n^\l \pi_{\m^\prime}\{\pi_\l,n^{\m^\prime}\}=\pi_\l \{n^\l,\pi_{n^\prime}\}+\pi_{\m^\prime} \{\pi_n, n^{\m^\prime}\}= \nonumber\\
&-\pi_\l\xi^{\l i}\pd_i \d(x-x^\prime) +\pi_{\m^\prime}\xi^{\m^\prime i} \pd_i \d(x-x^\prime)
 \eea
We have presented Dirac's results with (perhaps excessive) detail to highlight the generality and beauty of Dirac's algebra. Notice that  no dynamics enters into the proof; all results are purely kimematical as a consequence of having  assumed from the very beginning that there is no preferred notion of time. This is an eccentric luxury in flat spacetime, but it will become compulsory in General Relativity.

\subsection{The Arnowitt-Deser-Misner (ADM) formalism.}
Let us apply Dirac's ideas to the gravitational field. We shall assume that there is a foliation as before.
Remember the components of the spacetime metric in terms of the lapse and shift functions.
\bea
&&g_{00}= N^2\nonumber\\
&&g_{0i}=h_{ij}N^j\nonumber\\
&&g_{ij}=h_{ij}
\eea
whose inverse reads
\bea
&&g^{00}= N^{-2}\nonumber\\
&&g^{0 i}=-{N^i\over N^2}\nonumber\\
&&g^{ij}= h^{ij}+ {N^i N^j\over N^2}
\eea
Let us denote by $D_i$ the covariant derivative with respect to  the three-dimensional Levi-Civita connection associated to the induced metric, $h_{ij}$.
\par
It can be easily checked that
\be
D_j A_i=\nabla_\b A_\a \,\xi^\a_i \,\xi^\b_j
\ee

From the definition itself of the induced metric \cite{Eisenhart} it follows that
\be
\pd_\r g_{\a\b}D_k \s^\r \pd_i \s^\a \pd_j \s^b+ g_{\a\b} D_k (\pd_i \s^\a)\pd_j \s^\b+g_{\a\b}\pd_i \s^\a D_k (\pd_j \s^\b)=0
\ee
Peform now c yclic permutations in the above
\bea
&\pd_\r g_{\a\b}D_j \s^\r \pd_k \s^\a \pd_i \s^\b+ g_{\a\b} D_j (\pd_k \s^\a)\pd_i \s^\b+g_{\a\b}\pd_k \s^\a D_j (\pd_i \s^\b)=0\nonumber\\
&\pd_\r g_{\a\b}D_i \s^\r \pd_j \s^\a \pd_k \s^\b+ g_{\a\b} D_i (\pd_j \s^\a)\pd_i \s^\b+g_{\a\b}\pd_j \s^\a D_i (\pd_k \s^\b)=0\nonumber
\eea
Adding the first permutation to the second and subtracting  the third yields
\bea
&&0= g_{\a\b} D_j D_k \s^\a D_i \s^\b + D_k \s^\r D_i \s^\a D_j \s^\b {1\over 2}\left(\pd_\r g_{\a\b}+\pd_\b g_{\r\a}-\pd_\a g_{\b\r}\right)=\nonumber\\
&&g_{\a\b} D_j D_k \s^\a D_i \s^\b + D_k \s^\r D_i \s^\a D_j \s^\b \lbrace \a,\b\r\rbrace=\nonumber\\
&&=g_{\a\b} D_i \s^\b \left(D_k D_j \s^\a +\lbrace\,^\a_{\b\r}\rbrace D_j \s^\b D_k \s^\r\right)
\eea
This means that
\be
D_k D_j \s^\a =-\lbrace\,^\a_{\b\r}\rbrace D_j \s^\b D_k \s^\r+K_{jk} n^\a
\ee
where the normal component reads
\be
K_{jk}\equiv n_\a\left(D_k D_j \s^\a +\lbrace\,^\a_{\b\r}\rbrace D_j \s^\b D_k \s^\r\right)
\ee
Taking the three-dimensional covariant derivative $D_j$ 
\be
0=D_j\left(g_{\a\b}\pd_i\s^\a n^\b\right)=D_j g_{\a\b}\pd_i\s^\a n^\b+ g_{\a\b}D_j D_i\s^\a n^\b+g_{\a\b}D_i\s^\a D_j n^\b
\ee
On the other hand,
\be
D_j g_{\a\b}=D_j \s^\n \pd_\n g_{\a\b}=D_j \s^\n \left(\lbrace \a \n; \b\rbrace+\lbrace \b \n ; \a\rbrace\right)
\ee
so that the quantity we have just defined
\bea
&K_{jk}=n_\a \lbrace \,^\a_{\b\r}\rbrace D_j\s^\b D_k\s^\r-g_{\a\b} D_j \s^\a D_k n^\b-n^\b D_k \s^\r\left(\lbrace \a \r; \b \rbrace+ \lbrace \b\r ; \a\rbrace\right)D_j \s^\a=\nonumber\\
&-g_{\a\b} D_j \s^\a D_k n^\b - n^\b D_k \s^\r \lbrace \b \r ; \a \rbrace D_j \s^\a=-\xi_i^\a \nabla_\r n_\a \xi^\r_j
\eea
This tensor is called the {\em extrinsic curvature}, and represents the four-dimensional covariant derivative of the normal vector, projected on the surface.
\par
Let us now relate the Riemann tensor on the hypersurface (computed with the induced metric) with the corresponding Riemann tensor of the spacetime manifold. Those are the famous Gauss-Codazzi equations, which we purport now to derive.
\par
 They were one of the pillars of Gauss' {\em theorema egregium},  which asserts that {\em If a curved surface is developed upon any other surface whatever the measure of curvature in each point remains unchanged}.
\par

We start with
\bea\label{un}
&&0=D_j\left(g_{\a\b} n^\a n^\b\right)=D_j \s^\r\left(\lbrace \a\r;\b\rbrace+\lbrace\r\b;\a\rbrace\right)n^\a n^\b+ g_{\a\b}D_j n^\a n^\b+g_{\a\b} n^\a D_j n^\b=\nonumber\\
&&g_{\a\b}n^\b \left(D_j n^\a+\lbrace\,^\a_{\m\n}\rbrace D_j\s^{(\m}n^{\n)}\right)=g_{\a\b} n^\b \nabla_\m n^\a D_j \s^\m=n_\a \nabla_\m n^\a \xi_j^\m
\eea
On the other hand, the explicit expression for the extrinsic curvature reads
\be
K_{ij}=-\xi_i^\a \nabla_\r n_\a \xi^\r_j
\ee
First of all let us derive some properties of the extrinsic curvature. It is symmetric, $K_{ij}= K_{ji}$.
\be
-K_{ij}=\nabla_\b n_\a \xi^\a_i \xi^\b_j=-n_\a \nabla_\b \xi_i^\a \xi^\b_j
\ee
But remembering that
\be
\left[\xi^\b_j\pd_\b ,\xi^\a_i \pd_\a\right]=0
\ee
it follows
\be
-K_{ij}=-n_\a \,\xi_i^\a\,\nabla_\b\, \xi^\a_j=\nabla_\b\, n_\a\, \xi^\b_i\,\xi^\a_l=K_{ji}
\ee
This symmetry implies a very useful formula for the extrinsic curvature, namely
\be
-K_{ij}=\nabla_{(\b} n_{\a)}\, \xi^\a_i\, \xi^\b_j=\pounds(n)\,g_{\a\b}\, \xi^\a_i\,\xi^\b_j
\ee
By the way, in the physics jargon when $K_{ij}=0$ it is said that it is  a {\em moment of time symmetry}.

On the other hand, remembering that
\be
\xi_i^\a\xi^i_\b=g^\a_\b-n^\a n_\b
\ee
we deduce 
\be
-K_{ij}\xi^i_{\m}=-\left(g^\a_\m-n^\a n_\m\right)\nabla_\r n_\a \xi^\r_j=-\nabla_\r n_\m \xi^\r_j
\ee
(because of [\ref{un}]).
\par
Let us analyze the definition of extrinsic curvature in even more detail. We follow the explicit computations in the classic reference \cite{Eisenhart}
\bea
&&\left(D_k D_j D_i-D_j D_k D_i\right)\s^\a=\xi^\a_m h^{mh}R_{hijk}=D_k\left(-\lbrace\,^\a_{\b\r}\rbrace\xi^\b_i\xi^\r_j+K_{ij}n^\a\right)-\nonumber\\
&&-D_j\left(-\lbrace\,^\a_{\b\r} \rbrace \xi_i^\b \xi_k^\r+K_{ik}n^\a\right)=\pd_k \lbrace\,^\a_{\b\r}\rbrace\xi^\b_i\xi^\r_j-\lbrace\,^\a_{\b\r}\rbrace D_k\xi^\b_i\xi^\r_j-\lbrace\,^\a_{\b\r} \rbrace \xi_i^\b D_k\xi^\r_j+\nonumber\\
&&D_k K_{ij} n^\a + K_{ij} D_k n^\a+\pd_j \lbrace\,^\a_{\b\r}\rbrace \xi_i^\b\xi^\r_k-\lbrace\,^\a_{\b\r}\rbrace D_j\xi^\b_i \xi^\r_k+\lbrace\,^\a_{\b\r}\xi_i^\b D_j\xi^\r_k-D_j K_{ik}-K_{ik} D_j n^\a\nonumber
\eea
and using again the definition of the extrinsic curvature to eliminate the term with two derivatives,
\bea
&&\xi_m^\a h^{mr} R_{rijk}[h]=-\pd_k\lbrace\,^\a_{\b\r}\rbrace \xi^\b_i\xi^\r_j-\lbrace \,^\a_{\b\r}\rbrace \xi^\r_j\left(-\lbrace\,^\b_{\m\n}\rbrace \xi^\m_i\xi^\n_k + K_{ik} n^\b\right)+D_k K_{ij} n^\a + K_{ij} D_k n^\a+\nonumber\\
&&\pd_j\lbrace\,^\a_{\b\r}\rbrace\xi^\b_i\xi^\r_k+\lbrace\,^\a_{\b\r}\rbrace \xi_k^\r\left(-\lbrace\,^\b_{\m\n}\rbrace \xi^\m_i\xi^\n_j+K_{ij}n^\b\right)-D_j K_{ik} n^\a -K_{ik} D_j n^\a=\nonumber\\
&&n^\a\left( D_k K_{ij}-D_j K_{ik}\right)+K_{ij}\left(D_k n^\a +\lbrace\,^\a_{\b\r}\rbrace n^\b \xi^\r_k\right)-K_{ik}\left(D_j n^\a+\lbrace\,^\a_{\b\r} n^\b \xi_j^\r\right)-\nonumber\\
&&-\xi_i^\b \xi^\r_j\xi^\s_k\left(\pd_\s\lbrace\,^\a_{\b\r}\rbrace-\pd_\r \lbrace\,^\a_{\b\s}\rbrace-\lbrace\,^\a_{\l\r}\rbrace\lbrace\,^\l_{\b\s}+\lbrace\,^\a_{\l\s}\rbrace\lbrace\,^\l_{\b\r}\right)
\eea
Using again the definition of the extrinsic curvature, as well as the one of the full Riemann tensor, we get
\bea
&&\xi_m^\a h^{mr}\left(R_{rijk}[h]+K_{ij}K_{rk}-K_{ik}K_{rj}\right)-n^\a\left(D_k K_{ij}-D_j K_{ik}\right)=-\xi^\b_i\xi^\r_j\xi_k^\s R^\a\,_{\b\s\r}[g]\nonumber
\eea
This projects into the famous Gauss-Codazzi equations
\bea
&&R_{lijk}[h]+K_{il}K_{jk}-K_{ik}K_{lj}=\xi^\a_l\xi^\b_i\xi^\r_j\xi^\s_k R_{\a\b\r\s}[g]
\eea
This equation is telling us that the Riemann tensor associated to the induced metric is the total  tangent projection of the full four-dimensional Riemann tensor plus a couple of terms involving the extrinsic curvature.

It is also the case that when only one of the components of the four-dimensional Riemann tensor is projected along the normal, and all the others are tangent, then
\bea
&D_j K_{ik}-D_k K_{ij}=-n^\a \xi^\b_i\xi^\r_j\xi^\s_k R_{\a\b\s\r}[g]
\eea
Please note that not all components of the full Riemann tensor can be recovered from the knowledge of the Riemann tensor computed on the hypersurface plus the extrinsic curvature.
As a matter of fact, our main object of interest, which is the scalar of curvature (which we need for the Einstein-Hilbert (EH) action)
\be
\,^{(4)}R=\,^{(4)}R^{ij}\,_{ij}+ 2 \,^{(4)}R^i\,_{n i n}=^{(3)}R+ K^2-K_{ij}k^{ij}+ 2 \,^{(4)}R^i\,_{n i n}
\ee
This means that an explicit computation of $\,^{(4)}R^i\,_{n i n}$ is needed before the Einstein-Hilbert term could be written in the 1+(n-1) decomposition. To do that, consider the four-dimensional Ricci's identity
\be
\nabla_\g\nabla_\b n_\a-\nabla_\b\nabla_\g n_\a=R^\r\,_{\a\b\g} n_\r
\ee
Now
\bea
&&n^\b\left(\nabla_\g\nabla_\b n^\g-\nabla_\b\nabla_\g n^\g\right)=n^\b g^{\a\g} R^\r\,_{\a\b\g} n^\r\equiv R^{n\a}\,_{n\a}
\eea
Besides,
\bea
&&\nabla_\g n^\b \nabla_\b n^\g=\nabla_\g n_\b\left(n^\b n^\m+\xi^\b_i\xi^{\m i}\right)\left(n^\g n^\n +\xi_j^\g \xi^{j\n}\right)\nabla_\m n_\n=\nonumber\\
&&\nabla_\g n_\b\xi^\b_i\xi^{\m i}\xi_j^\g \xi^{j\n}\nabla_\m n_\n=-K_{ij}K^{ij}
\eea
Summarizing,
\bea
&R^{n\a}\,_{n\a}=n^\b\nabla_\g\nabla_\b n^\g-n^\b \nabla_\b \nabla_\g n^\g=\nabla_\g\left(n^\b\nabla_\b n^\g\right)-\nabla_\g n^\b \nabla_\b n^\g-\nabla_\b\left(n^\b \nabla_\g n^\g\right)+\nonumber\\
&+\nabla_\b n^\b \nabla_\g n^\g=\nonumber\\
&=\nabla_\g\left(n^\b\nabla_\b n^\g -n^\g \nabla_\b n^\b\right)+K_{ij}K^{ij}-K^2
\eea
The determinants are related through
\be
\sqrt{\,^{(4)}~g}=N~\sqrt{\,^{(3)}~g}
\ee
The EH lagrangian can then be written as follows
\be
L_{EH}=N~\sqrt{\,^{(3)}g}\left(\,^{(3)}R+K_{ij}K^{ij}-K^2\right)-\pd_\a V^\a\equiv L^\prime_{EH}-\pd_\a V^\a
\ee
where
\be
V^\a\equiv 2\sqrt{\,^{(4)}g}\left(n^\b\nabla_\b n^\g -n^\g \nabla_\b n^\b\right)
\ee
The resulting lagrangian, $L^\prime_{EH}$ does not contain $\dot{N}$ or $\dot{N}^i$, and does contain only first time derivatives of $g_{ij}$. This lagrangian differs from  the EH one by a total derivative. This is irrelevant  for the EM, but it has importance whenever the spacetime manifold has got a boundary.
\par
At any rate, this is the starting point of the ADM hamiltonian formalism.
There are the primary constraints
\be
\pi^\m={\d L\over \d\dot{N}^\m}=0
\ee
In order to compute the spacelike momenta, consider 
\bea
&\dot{h}_{ij}=\pounds(N^\a) h_{ij}=\xi^\a_i \xi^\b_j \pounds(N^\a) g_{\a\b} =\nonumber\\
&\text{(Remembering that this Lie derivative of the spacelike basis vectors vanishes)}=\nonumber\\
&=\xi^\a_i \xi^\b_j\left(\nabla_\a N_\b+\nabla_\b N_\a\right)=\xi^\a_i \xi^\b_j\left(\nabla_\a \left(N n_\b +{\cal N}_\b\right)+\nabla_\b \left(N n_\a+{\cal N}_\a\right)\right)=\nonumber\\
&=2 N K_{ij}+D_i N_j+D_j N_i
\eea

Then
\be
{\d K_{ij}\over \d \dot{h}_{kl}}={1\over 4 N}\left(\d^k_i\d^l_j+\d^k_j\d^l_i\right)
\ee
and
\be
{\d K\over \d \dot{k}_{kl}}={1\over 2 N} h^{kl}
\ee
and
\be
\pi^{ij}=\sqrt{h}\left(K^{ij}- K h^{ij}\right)
\ee
Let us compute now the hamiltonian
\be
H\equiv \int d^3 x \left(\pi_\m \dot{N}^\m+\pi_{ij}\dot{h}^{ij}\right)-L
\ee 
where
\be
L=N\sqrt{|h|} \left( R[h]+K_{ij}K^{ij}-K^2\right)
\ee
Now, it is clear that
\be
\sqrt{h}\left(K_{ij}K^{ij}-K^2\right)={1\over \sqrt{h}}\left(\pi_{ij}\pi^{ij}-{\pi^2\over 2}\right)
\ee

\par
We just derived

\be
\dot{h}_{ij}=2 N K_{ij}+D_i N_j+D_j N_i
\ee

Summarizing,
\bea
&&H=\pi_{ij} \dot{h}^{ij}-L=\nonumber\\
&&=\pi^{ij}\left({2 N\over \sqrt{|h|}}\left(\pi_{ij}-{1\over 2}\pi h_{ij}\right)+D_i N_j+D_j N_i\right)-N\left(\sqrt{|h|} R[h]+{1\over \sqrt{|h|}}\left(\pi_{ij}\pi^{ij}-{1\over 2}\pi^2\right)\right)=\nonumber\\
&&=\int d^3 x \left( N {\cal H}+N^i{\cal H}_i\right)
\eea
where (dropping surface terms)
\bea
&&{\cal H}={1\over 2 \sqrt{h}}\left(h_{ik} h_{jl}+h_{il}h_{jk}-h_{ij} h_{kl}\right)\pi^{ik}\pi^{kl}-\sqrt{h} R[h]={1\over \sqrt{h}}\left(\pi_{ij}\pi^{ij}-{1\over 2}\pi^2\right)-\sqrt{h} R[h]\nonumber\\
&&{\cal H}_i\equiv -2 D_j \pi_i^j
\eea
Now we have the following constraints
\bea
&&\pi_\m\sim 0\nonumber\\
&&N^\m-C^\m\sim 0\nonumber\\
&&{\cal H}\sim 0\nonumber\\
&&{\cal H}_i\sim 0
\eea
and the corresponding brackets
\bea
&&\{\pi^\m,N^\r-C^\r\}\sim g^{\m\r}\d(x-x^\prime)\nonumber\\
&&\{\pi_\m,{\cal H}\}\sim 0\nonumber\\
&&\{\pi_\m,{\cal H}_i\}\sim 0\nonumber\\
&&\{N^\m-C^\m,{\cal H}\}\sim 0\nonumber\\
&&\{N^\m-C^\m,{\cal H}_i\}\sim 0
\eea
The other brackets got the universal Dirac-Schwinger form, which is valid for any diffeomorphism invariant field theory
\bea
&&\{{\cal H}(x),{\cal H}(x^\prime)=\left({\cal H}^i(x)+{\cal H}^i(x^\prime)\right)\pd_i\d(x-x^\prime)\}\sim\nonumber\\
&&\{{\cal H}_i(x),{\cal H}(x^\prime)\}\sim{\cal H}(x)\pd_i \d(x-x^\prime)\nonumber\\
&&\{{\cal H}_i(x),{\cal H}_j(x^\prime)\}\sim{\cal H}_i(x^\prime)\pd_j \d(x-x^\prime)+{\cal H}_j(x)\pd_i\d(x-x^\prime)
\eea

\subsection{Careful analysis of the Boundary terms}
The purpose of this section is to give a detailed treatment of boundary terms following Brown and York \cite{Brown} (confer also the careful treatment in \cite{Poisson}).
 
 \begin{center}
\includegraphics[width=0.49\linewidth]{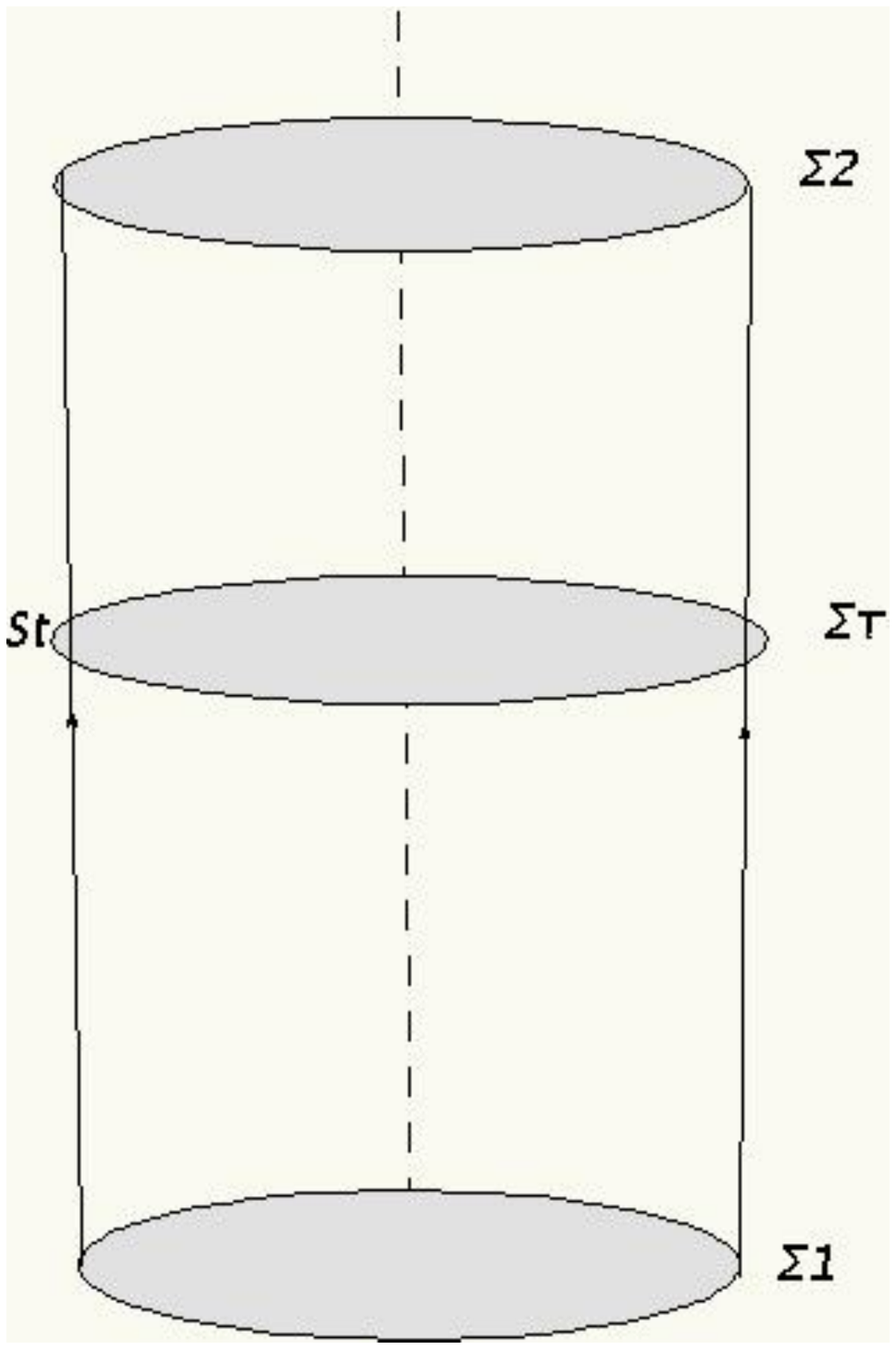}
\captionof{figure}{The spacetime domain under consideration, with the boundaries specified.}
\end{center}

 Consider a tubular domain $D$ of spacetime, whose boundary has three different pieces: The two caps at the initial and final times, $\Sigma_{t_1}$ and $\Sigma_{t_2}$. Those are spacelike, codimension one hypersurfaces (that is $d=n-1$). The physical space-time of course has dimension $n=4$, but the analysis can easily be made for general dimension $n$. Then there is the "boundary at infinity", $r=R\rightarrow\infty $, which is the surface of a cylinder, also of codimension one, but timelike instead of spacelike. We shall call it $B\equiv \pd D$. Now this boundary can be understood as generated by the union of all the codimension two boundaries of the constant time hypersurfaces
 \be
\left( B= \pd D \right)\equiv \quad\cup_t\quad\left( S_t=\pd\Sigma_t\right)
 \ee
 \bi
 \item
 An intuitive grasp of the general situation can stem from the trivial example in flat space, to which we are going to refer all the time.
 \be
 D\equiv \{r\leq R \quad t_1\leq t\leq  t_2\}
 \ee
 In this way the caps are defined by the solid balls
 \be
 \Sigma_t\equiv \{r\leq R \quad\cup\quad t=\text{constant}\}
 \ee
 The embedding in spacetime is simply
 \bea
&&y^0= t\nonumber\\
&& y^i=x^i
\eea
so that the induced tangent vectors is
\be
\xi^\a_i\equiv {\pd y^\a\over \pd x^i}=\begin{pmatrix}0&0&0\\1&0&0\\0&1&0\\0&0&1\end{pmatrix}
\ee 
and the normal vector
\be
n^\a\equiv \left( 1,0,0,0\right)
\ee
The induced metric reads
\be
h_{ij}\equiv \eta_{\a\b}\xi^\a_i \xi^\b_j=-\d_{ij}
\ee
The normal to $\Sigma_t$ in  Minkowski space is
\be
n^\a=\left(1,0,0,0\right)
\ee
 The  boundary of such caps are the two-spheres
 \be
 S_t\equiv \{r=R\quad\cup\quad t=\text{constant} \}
 \ee
 We can choose polar coordinates $\theta_a\equiv (\theta,\phi)$. The imbedding matrix of the boundary in $\Sigma_t$ using these is
 \be
 \xi^i_a\equiv {\pd y^i\over\pd \theta^a}=\begin{pmatrix}\text{cos}~\theta~\text{cos}~\phi&-\text{sin}~\theta~\text{sin}~\phi\\\text{cos}~\theta~\text{sin}~\phi&\text{sin}~\theta~\text{cos}~\phi\\-\text{sin}~\theta&0\end{pmatrix}
 \ee
 It is equivalent to use
 \bea
 &&\theta_1\equiv {y_1\over R}\nonumber\\
 &&\theta_2\equiv {y_2 \over R}
 \eea
 then
 \be
 y_3\equiv \theta_3 R=R\sqrt{1-\theta_1^2-\theta_2^2}
 \ee
 The embedding matrix is now
 \be
 \xi^i_a=\begin{pmatrix}R&0\\0&R\\-R{\theta_1\over \sqrt{1-\theta_1^2-\theta_2^2}}&-R{\theta_2\over \sqrt{1-\theta_1^2-\theta_2^2}}\end{pmatrix}
 \ee
 The induced metric reads
 \be
 \s_{ab}\equiv \xi^i_a h_{ij} \xi^j_b=-\vec{\xi}_a.\vec{\xi}_b=-{R^2\over 1-\theta_1^2-\theta_2^2}\begin{pmatrix}1-\theta_2^2&\theta_1\theta_2\\\theta_1\theta_2&1-\theta_1^2\end{pmatrix}
 \ee
 
 The contravariant metric reads
 \be
 \s^{ab}=-{1\over R^2}\begin{pmatrix}1-\theta_1^2&-\theta_1\theta_2\\-\theta_1\theta_2&1-\theta_2^2\end{pmatrix}
 \ee
  Out of the two embedding matrices we can draw the composition
 \be
 \xi^\a_a\equiv e^\a_j e^j_a\equiv\begin{pmatrix}0&0\\R&0\\0&R\\-R{\theta_1\over \sqrt{1-\theta_1^2-\theta_2^2}}&-R{\theta_2\over \sqrt{1-\theta_1^2-\theta_2^2}}\end{pmatrix}=\begin{pmatrix}0&0\\\text{cos}~\theta~\text{cos}~\phi&-\text{sin}~\theta~\text{sin}~\phi\\\text{cos}~\theta~\text{sin}~\phi&\text{sin}~\theta~\text{cos}~\phi\\-\text{sin}~\theta&0\end{pmatrix}
 \ee
 
 The normal to the boundary in $\Sigma_t$ is
 \be
 \n^i=(\text{sin}~\theta~\text{cos}~\phi,\text{sin}~\theta~\text{sin}~\phi,\text{cos}~\theta)=\left({x\over R},{y\over R},{z\over R}\right)=\left(\theta_1,\theta_2,\sqrt{1-\theta_1^2-\theta_2^2}\right)
 \ee
 The extrinsic curvature of $S_t\hookrightarrow \Sigma_t$ reads
 \be
 k_{ab}\equiv \nabla_j \n_i \xi^j_a \xi^i_b={1\over R}~\d_{ij}~\xi^j_a \xi^i_b=-\s_{ab}
 \ee
 Let us now examine the constructs
 \be
 \n^i \n^j e_i^\a e_j^\b=\begin{pmatrix}0&0&0&0\\0&\theta_1^2&\theta_1\theta_2&\theta_1\sqrt{1-\theta_1^2-\theta_2^2}\\0&\theta_2\theta_1&\theta_2^2&\theta_2\sqrt{1-\theta_1^2-\theta_2^2}\\0&\theta_1\sqrt{1-\theta_1^2-\theta_2^2}&\theta_2\sqrt{1-\theta_1^2-\theta_2^2}&1-\theta_1^2-\theta_2^2\end{pmatrix}
 \ee
 \be
 \s^{ab}e^\a_a e^\b_b=-R^2\begin{pmatrix}0&0&0&0\\0&1-\theta_1^2&-\theta_1\theta_2&-\theta_1\sqrt{1-\theta_1^2-\theta_2^2}\\0&-\theta_1\theta_2&1-\theta_2^2&-\theta_2\sqrt{1-\theta_1^2-\theta_2^2}\\0&-\theta_1\sqrt{1-\theta_1^2-\theta_2^2}&-\theta_2\sqrt{1-\theta_1^2-\theta_2^2}&\theta_1^2+\theta_2^2\end{pmatrix}
 \ee
All this explicitly checks that
 \be
 n^\a n^\b-\n^\a \n^\b+\s^{ab}e^\a_a e^\b_b=\eta^{\a\b}
 \ee
 
 The timelike boundary is just $S_2\times \mathbb{R}$
 \be
 B\equiv \cup_t S_t
 \ee
 Its three coordinates are just $x^m=(t,y_1,y_2)$ ( they could equally well be chosen as $(t,\theta,\phi)$ or even $(t,\theta_1,\theta_2$).
 The embedding matrix reads
 \be
 \xi^\a_m\equiv R\begin{pmatrix}1&0&0\\0&1&0\\0&0&1\\0&-{\theta_1\over\theta_3}&-{\theta_2\over\theta_3}\end{pmatrix}
 \ee
 so that the induced metric is just
 \be
 ds^2= dt^2-{R^2\over \theta_3^2}\left((1-\theta_2^2)d\theta_1^2+2\theta_1\theta_2 d\theta_1d\theta_2 +(1-\theta_1^2)d\theta_2^2\right)
 \ee
 The normal vector is
 \be
 n^\a={1\over R}\left(0,x,y,z\right)
 \ee
 so that the extrinsic curvature of $B\hookrightarrow M$ reads
 \be
 \kappa_{mn}\equiv\nabla_\a n_\b \xi^\a_m \xi^\b_n=\begin{pmatrix}0&0\\0&\s_{ab}\end{pmatrix}
 \ee

\item
Let us now draw from the example to the general case. 
The surfaces $S^{n-2}_t\equiv\pd \Sigma^{n-1}_t$ provide a foliation of the timelike boundary $B_{n-1}\hookrightarrow V_n$ of the domain of spacetime under consideration. The coordinates in $S^{(n-2)}_t$ will be denoted by $\theta_a\quad a=1\ldots n-2$. The imbedding $S_{n-2}\hookrightarrow \Sigma_{n-1}$ is described by
\be
\theta\in S_{n-2} \hookrightarrow ~y^i(\theta^a)\in \Sigma_{n-1}\quad (i=1\ldots n-1)\quad (a=1\ldots n-2)
\ee
The imbedding of $S$ in $\Sigma$ defines in a natural way $(n-2)$ tangent space vectors
\be
 \xi^i_a\equiv {\pd y^i\over \pd \theta^a} 
\ee
The unit normal to $S_{n-2}$ in $\Sigma_{n-1}$ will be denoted by $\n^i$, and out of it we construct a vector
\be
\n^\a\equiv \n^i \xi_i^\a\in T(S)
\ee
which is such that it is unitary $\n.\n=1$ and is tangent to $\Sigma_{n-1}$, that is, $\n.n=0$. 
There are also $n-2$ spacetime vectors obtained by combining the two embeddings $S\hookrightarrow \Sigma$ and $\Sigma\hookrightarrow M$:
\be
\xi^\a_a\equiv \xi^\a_i \xi^i_a
\ee
The induced metric  in $S_{n-2}\equiv \pd \Sigma_{n-1}$ is
\be
ds^2\equiv\sum_{a,b=1}^{n-2} \s_{ab} d\theta^a d\theta^b = h_{ij}\xi^i_a \xi^j_b d\theta^a d\theta^b=g_{\a\b}e^\a_a e^\b_b d\theta^a d\theta^b
\ee
The spacetime metric can be recovered from 
\be
g^{\a\b}=-\n^\a\n^\b+n^\a n^\b+\s^{ab}e_a^\a e_b^\b
\ee
The extrinsic curvature of $S_{n-2}\hookrightarrow \Sigma_{n-1}$ is defined as usual
\be
k_{ab}\equiv \nabla_j \n_i \xi^j_a \xi^i_b
\ee
It is possible to choose the coordinates $\theta^a$ in such a way that they intersect $S^{n-2}_t\equiv \pd \Sigma^{n-1}_t$ orthogonally.

This means that the vector $n^\a$ is the tangent vector to the timelike flow
\be
N n^\a=\left({\pd x^\a\over \pd t}\right)_\theta
\ee

The set of all $S_t^{n-1}$ for varying t do foliate the timelike boundary of spacetime $B_{n-1}\equiv \pd V_n$. In this boundary $B_{n-1}$ we can also introduce coordinates $z^m\quad m=1\ldots n-1$ (one of which is timelike), and the corresponding $(n-1)$ vectors
\be
\xi^\a_m\equiv{\pd x^\a\over \pd z^m}
\ee
The induced metric is
\be
\g_{mn}=g_{\a\b} \xi^\a_m \xi^\b_n
\ee
and we can write the {\em completeness relation}
\be
g_{\a\b}=-\n_\a\n_\b+\g_{mn}\xi^m_\a \xi^n_\b
\ee
It is simplest to choose (as we did in our explicit example)
\be
z^m\equiv\left(t,\theta^a\right)
\ee
then
\be
dx^\a=\left({\pd x^\a\over \pd t}\right)_\theta dt + \left({\pd x^\a\over \pd \theta^a}\right)_t d\theta^a= N n^\a dt+ \xi^\a_a d\theta^a
\ee
in such a way that
\be
\left.ds^2\right|_B=\g_{mn} dz^m dz^n=  N^2 dt^2+\s_{ab} d\theta^a d\theta^b
\ee
and the determinant obeys
\be
|\g|=N^2 \s
\ee
Finally, the extrinsic curvature of $B_{n-1}\hookrightarrow V_n$ is
\be
\kappa_{mn}=\nabla_\b \n_\a \xi^\a_m \xi^\b_n
\ee
\item 
Let us apply all this mathematics to the Einstein-Hilbert action. We consider a tubular region of the full spacetime bounded by two spacelike hypersurfaces of constant time, $\Sigma_2$ and $\Sigma_1$, and the surface of the asymptotic cylinder, $B$
\be
\pd V_n=\Sigma_2-\Sigma_1+B
\ee
This is the generalization to an arbitrary spacetime of the construction made in the example.
\par
The full EH action, including the boundary term as well as the total derivative neglected when constructing the generic hamiltonian is given by
\bea
&&S_{EH}={c^3\over 16\pi G}\int_{t_1}^{t_2} dt \int_{\Sigma_t} N\sqrt{|h|} d^{n-1}y\left( R[h]+K_{ij}K^{ij}-K^2-\right.\nonumber\\
&&\left.-2 \nabla_\a\left(n^\m\nabla_\m n^\a-n^\a\nabla_\l n^\l\right)\right)+{1\over 8\pi G}\int_{\Sigma_{t_1}} K-{1\over 8\pi G}\int_{\Sigma_{t_2}} K-{1\over 8\pi G}\int_{B} K\nonumber
\eea
Perhaps we should comment at this point that the boundary term, which is precisely proportional to the boundary integral of the trace of the extrinsic curvature, was first introduced by York and later on used by Gibbons and Hawking, in order for the variational principle to be well defined when the region of integration has a boundary. It is indeed compulsory from the present viewpoint, as we shall witness in a moment.
\par

The  total derivative piece in the expansion of $R$ which yields a boundary piece
\be
-2\int_{\pd V}\left(n^\b\nabla_\b n^\a-n^\a\nabla_\b n^\b\right) n_\a \sqrt{|h|}d^{n-1}y=-2\int K\sqrt{|h|} d^{n-1} y
\ee
This precisely cancel the boundary term in the action coming from $\Sigma_t$. The only surviving contribution comes from the timelike boundary, $B$, that is
\be
-2\int_B\left(n^\b\nabla_\b n^\a-n^\a\nabla_\b n^\b\right)\n_\a\sqrt{|\g|}d^{n-1}z= 2 \int_B \nabla_\b \n_\a n^\a n^\b \sqrt{|\g|}d^{n-1}z
\ee
Summarizing
\be
{16\pi G\over c^3} S_{EH}=\int_{t_1}^{t_2}+2\int_B\left(\kappa+\nabla_\b\n_\a n^\a n^\b\right)\sqrt{|\g|} d^{n-1}z
\ee
Let us use now the fact that the timelike boundary  $B$ is foliated by $S_t$
\be
\kappa=\g^{ij}\kappa_{ij}=\g^{ij}\nabla_\b\n_\a e^\a_i e^\b_j=\nabla_\b \n_\a\left(g^{\a\b}-\n^\a\n^\b\right)
\ee
This means that 
\be
 \kappa+ \nabla_\b\n_\a n^\a n^\b=\nabla_\b\n_\a\left(g^{\a\b}-\n^\a\n^\b+n^\a n^\b\right)=\nabla_\b\n_\a e^\a_a e^\b_b=\s^{ab}\kappa_{ab}=k
\ee
so that 
\be
 \int_B=2 \int_{S^{(n-2)}_t} k N \sqrt{|\s|} d^{n-2}\theta
\ee
As was already clear from the explicit example, this integral diverges even $R\rightarrow\infty$ even in flat space. In order to refer all expressions to this value, so that the action in flat space vanishes, it is often subtracted a term in the action
\be
\Delta D\equiv -{2\over 16\pi G}\int_B  k_0 N 
\ee
where $k_0$ represents the extrinsic curvature of $S_{n-2}$ embedded in flat space.

\par
The boundary terms in the hamiltonian read
\be
H_{\text{boundary}}=-2{1\over 16\pi G}\int_{S_{n-2}} \left(N\left(k-k_0\right)-N_i\left(K^{ij}- K h^{ij}\right)\n_j\right) \sqrt{|s|}~d^{n-2}~\theta
\ee
(where $K_{ij}$ is to be understood as a functional of the hamiltonian variables $h_{ij}$ and $\pi_{kl}$. To be specific,

\be
K^{ij}\equiv{16\pi G\over\sqrt{|h|}}\left(\pi^{ij}-{1\over 2}\pi h^{ij}\right)
\ee
This boundary term yields the value of the energy for the gravitational field. It depends of the foliation chosen as well as on the lapse and shift which are arbitrary.
When the space is asymptotically flat, representing flat asymptotic coordinates as $(T,X^i)$, it is possible to choose $\Sigma_t$ so that  goes into $T=\text{constant}$. It is clear that
\be
N^\a\rightarrow N\left({\pd x^\a \over \pd T}\right)+ N^i \left({\pd x^\a\over \pd X^i}\right)
\ee
It is then natural to define the {\em ADM mass} associated to a given solution by choosing a FIDO at rest at infinity, that is, $N=1$, $N^i=0$, so that

\be
N^\a\rightarrow \left({\pd x^\a \over \pd T}\right)
\ee
and the flow generates a time translation at infinity.
Then 
\be
M\equiv -\lim_{R\rightarrow\infty}{1\over 8\pi G}\int_{S_{n-2}}\left(k-k_0\right)\sqrt{|\s|} d^{n-2}\theta
\ee
\item 
As the simplest of all possible exercises, let us compute the ADM mass for the four-dimensional Schwarzschild's spacetime,
\be
ds^2= \left(1-{r_S\over r}\right)dt^2- {dr^2\over 1-{r_S\over r}}-r^2 d\Omega_2^2
\ee
Let us choose $\Sigma_t$ to be really the surfaces of constant Schwarzschild time. Then the unit normal is given by
\be
n^\a={1\over \sqrt{1-{r_S\over r}}}\left(1,0,0,0\right)
\ee
The induced metric in $\Sigma_t$ is
\be
h_{ij} dx^i dx^j={1\over 1-{r_S\over r}}dr^2+r^2 d\Omega_2^2
\ee
The boundary $S\equiv\pd \Sigma$  is again the two-sphere $r=R$, and the unit normal is
\be
\n=\sqrt{1-{r_S\over r}}{\pd\over \pd r}
\ee
The induced metric is
\be
\s_{ab}d\theta^a d\theta^b=R^2 d\Omega_2^2
\ee
The extrinsic curvature reads

\bea
&&k=\nabla_a n^a={1\over \sqrt{h}}\pd_i \left(\sqrt{|h|} \n^i\right)={\sqrt{1-{r_S\over r}}\over r^2 }\pd_r\left[{r^2\over \sqrt{1-{r_S\over r}}}\sqrt{1-{r_S\over r}}\right]=\nonumber\\
&&{2\over R}\sqrt{1-{r_S\over R}}
\eea
On the other hand
\be
k_0={1\over r^2}\pd_r r^2={2\over R}
\ee
It is then a fact that
\be
k-k_0\sim -{r_S\over R^2}
\ee

so that
\be
M_{ADM}={1\over 8\pi G}.4\pi r_S=M
\ee
This is actually the reason why we have defined $r_S\equiv 2 G M$.
\item
The ADM mass does not capture the mass loss due to radiation. In order to do that, it is necessary to choose the boundary at null infinity, instead of at spatial infinity. The corresponding mass is called the {\em Bondi mass} \cite{Bondi}
\be
M_{\text{bondi}}\equiv -{1\over 8\pi G}\int_{v\rightarrow\infty}\left(k-k_0\right)
\ee
where the retarded time has been defined as usual
\be
u\equiv t-r
\ee
and the advanced time
\be
v\equiv t+r
\ee
Let us work this out explicitly in an example \cite{Poisson} . Consider the source
\be
T_{\m\n}\equiv -{1\over 4\pi G  r^2}{d M(u)\over du}~l_\a l_\b
\ee
where now $u$ is the null Schwarzschild coordinate
\be
u\equiv t-r-r_S\, \log~\left({r\over r_S}-1\right)
\ee
and the mass (and also $r_S(u)\equiv 2 G M(u)$) depend on $u$. The null vector
\be
l \equiv\pd_u
\ee
The matter represented by the energy-momentum tensor as above is referred to as {\em null dust}. The solution of Einstein's equations is called the Vaidya metric and reads
\be
ds^2=\left(1- {2 G M(u)\over r}\right) du^2+2 du dr -r^2d\Omega^2
\ee
The contravariant metric in the sector $(u,r)$ reads
\be
g^{\m\n}=\begin{pmatrix}0&1\\1&-\left(1- {2 G M(u)\over r}\right)\end{pmatrix}
\ee
Let us consider again the surface $\Sigma_t$ where 
\be
u+r=\text{constant}
\ee
Its covariant normal reads
\be
n_\a\sim\left(1,1\right)
\ee
so that the normal vector
\be
n\sim\left(g^{uu}+g^{ur},g^{ru}+g^{rr}\right)=\left(1,{2 G M(u)\over r}\right)
\ee
Normalizing
\be
n={1\over \sqrt{1+{2GM(u)\over r}}}\left(\pd_u+{2 G M(u)\over r}\pd_r\right)
\ee
The induced metric in $\Sigma$ is obtained by substituting $du=-dr$, so that
\be
ds^2=-\left(1+{2 G M(u)\over r}\right)dr^2- r^2 d\Omega^2
\ee
The boundary $\pd \Sigma$ is just the sphere $r=R$. The normal is 
\be
\n\equiv {1\over \sqrt{1+{2 G M(u)\over r}}}{\pd\over\pd r}
\ee
The extrinsic curvature reads
\be
k\equiv\nabla_a \n^a={2\over R \sqrt{1+{2 G M(u)\over R}}}\sim {2\over R}\left(1-{G M(u)\over R}+\ldots\right)
\ee
The indiced metric on the boundary is just
\be
ds^2=-R^2 d\Omega^2
\ee
The extrinsic curvature of a surface of the same intrinsic geometry, only that embedded in flat space is
\be
k_0={1\over r^2}\pd_r(r^2)={2\over R}
\ee
so that 
\be
k-k_0=-{2 G M(u)\over R^2}
\ee
If we integrate now on spatial infinity $R\rightarrow\infty$, this means that we keep $t\equiv u + r$ constant, so that
$u\sim -R\rightarrow-\infty$. This means that
\be
M_{ADM}=M(u=-R)
\ee
If we integrate now on null infinity $v\rightarrow \infty$, while $u$ is kept fixed, then
\be
M_B= M(u)
\ee
the mass function.

\ei

\begin{center}
\includegraphics[width=0.49\linewidth]{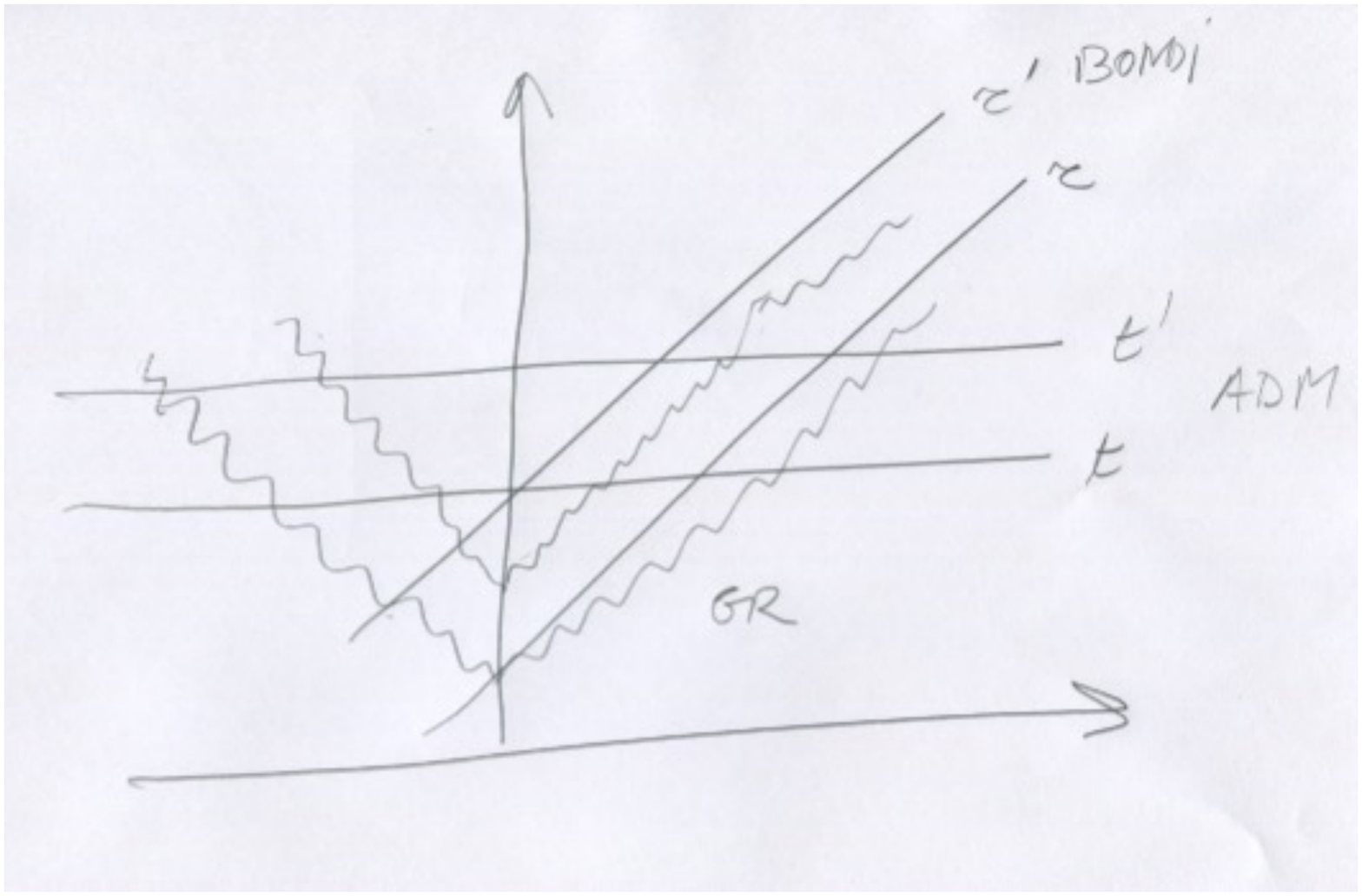}
\captionof{figure}{ADM energy does not measure the loss of energy due to gravitational radiation. In fact $ADM(t)=ADM(t^\prime)$. Bondi energy does it. $B(\t)\geq B(\t^\prime)$}
\end{center}

\subsection{Canonical quantum gravity}
The first attempt to quantize the gravitational field \cite{DeWitt} stems from the preceding canonical approach, just by converting the Poisson (or Dirac) brackets into commutators as in
\be
\left[\pi_{ij}(x),h^{kl}(\xp)\right]=-i \d_{ij}^{kl} \d(x,\xp)
\ee
There are lots of mathematical ambiguities of operator ordering, and also defining in a precise way products of distributions, and so on, and also physical problems, to which now turn. But still, the approach offers some glimpses of what a true quantum geometry theory might be. 
\par
From the physical point of view, it has been realized since long that this whole  approach suffers from the  {\em frozen time
problem}, i.e., the generic Hamiltonian reads

\be
H\equiv \int d^3 x\,\left(N{\cal H}+N^i{\cal H}^i\right)\nonumber
\ee
so that acting on physical states of the Hilbert space with the corresponding operator
\be
{\hat H}|\psi\rangle=0
\ee

in such a way that Schr\"odinger's equation 

\be
i\frac{\pd}{\pd t}|\psi\rangle={\hat{H}}|\psi\rangle
\ee
seemingly forbids any time dependence. There is no known way out of this dilemma. Some concept of time can be recovered however in the semiclassical approximation \cite{Banks}, although it is not clear how to connect with the minkowskian time..

\par
There are many unsolved problems in this approach, which has been kept at a formal level.
The first one is an obvious operator ordering ambiguity owing to the nonlinearity of the classical expression for the hamiltonian.
In the same vein, it is not clear whether it is possible to make the constraints hermitian. There is no clear candidate for a positive semi-definite scalar product
Besides, it is not clear that one recovers the full diffeomorphism invariance from the
Dirac- Schwinger algebra. Actually, it is not even known whether this is necessary; that is, 
what is the full symmetry of the quantum theory.
\par

We can proceed further, still formally\footnote{ It is bound to be formal as long as
the problem of the infinities is not fully addressed. We know from the analysis of this
representation for gauge theories in the lattice that those are the most difficult 
problems to solve.}, using the Schr\"odinger representation
defined in such a way that
\be
(\hat{h}_{ij}\psi)[h]\equiv h_{ij}(x)\psi [h]
\ee
and
\be
(\hat{\pi}^{ij}\psi)[h]\equiv - i \hbar \frac{\d\psi}{\d h_{ij}(x)}[h]
\ee
If we assume that diffeomorphisms act on wave functionals as:
\be
\psi[f^{*}h]=\psi[h]
\ee
then the whole setup for the quantum dynamics of the gravitational field lies in
Wheeler's {\em superspace} (nothing to do with supersymmetry) which is
the set of three-dimensional metrics modulo three-dimensional diffeomorphisms : 
$Riem(\Sigma)/Diff(\Sigma)$.
\par

The Hamiltonian constraint then implies the famous {\em Wheeler-DeWitt} equation.

\be
-\hbar^2 2\kappa^2 G_{ijkl}\frac{\d^2 \psi}{\d h_{ik}\d h_{jl}}[h]
-\frac{h}{2\kappa^2} R^{(3)}[h]\psi[h]=0
\ee
where the DeWitt tensor is:
\be
G_{ijkl}\equiv\frac{1}{\sqrt{h}}\bigg(h_{ij}h_{kl}-\frac{1}{2}h_{ik}h_{jl}\bigg)
\ee
Needless to say, this equation, suggestive as it is,  is plagued with ambiguities.
The manifold of positive definite metrics has been studied by DeWitt. He showed that it has 
signature $(-1,+1^5)$, where the timelike coordinate is given by the breathing mode
of the metric:
\be
\zeta=\sqrt{\frac{32}{3}}h^{1/4}
\ee
and in terms of other five coordinates $\zeta^a$ orthogonal to the timelike coordinate, 
the full metric reads
\be
ds^2=-d\zeta^2 + \frac{3}{32}\zeta^2 g_{ab}d\zeta^a d\zeta^b
\ee

with
\be
g_{ab}= tr\, h^{-1}\pd_a h h^{-1}\pd h
\ee
The  five dimensional submanifold with metric $g_{ab}$ is the coset space
\be
SL(3,\mathbb{R})/SO(3)
\ee
It has been much speculated whether the timelike character of the dilatations lies at the root
of the concept of time. The Wheeler-deWitt equation can be written in a form quite similar to
the Klein-Gordon equation:
\be
\bigg(-\frac{\pd^2}{\pd\zeta^2}+\frac{32}{3 \zeta^2}g^{ab}\pd_a\pd_b+\frac{3}{32}\zeta^2 R^{(3)}
\bigg)\Psi=0
\ee

The analogy goes further in the sense that also here there is a naturally defined scalar product
which is not positive definite:
\be
(\psi,\chi)\equiv\int_{\Sigma}\psi^{*}d\Sigma^{ij}G_{ijkl}\frac{\d \chi}{i\d h_{kl}}
-\chi^{*}d\Sigma^{ij}G_{ijkl}\frac{\d \psi}{i\d h_{kl}}
\ee

\par
There has been a lot of activity in canonical quantum gravity following the discovery by Ashtekar \cite{Ashtekar}  of a new set of variables cf. for example, Rovelli's book in \cite{Books}. 
\par
It is my opinion that despite some formal interest in many cases,  this approach fails to comply  with the correspondence principle, in the sense that is not connected smoothly with either classical general relativity or else perturbative quantum corrections. This does not mean that some concepts and techniques developed in this approach could not be useful in the quantum regime very far from the classical limit; we don't know yet.
\par
At any rate, we shy away from treating this approach further in this lectures and refer to the literature to the interested reader. For a thorough review of this viewpoint cf. Thiemann's book \cite{Thiemann}.

\subsection{The Hartle-Hawking state}
Let us first review Vilenkin's idea of creation of universes out of nothing \cite{Vilenkin}. Consider  a charged particle of mass $m$ and charge $q$ in a constant electric field $E$ moving in the $(t,x)$ plane. Its trajectory is given by the hyperbola
\be
(t-t_0)^2-(x-x_0)^2=-R^2
\ee
where $R\equiv \left|{m \over q E}\right|$. When
\be
x-x_0=\pm R
\ee
there is a turning point at which $t=t_0$; the solution does not exist for 
\be
x-x_0 <R
\ee
The euclidean trajectory (a compact instanton or bounce) is obtained by the change 
\be
t\rightarrow i \t
\ee
and is just a circle
\be
(\t-\t_0)^2+(x-x_0)^2=R^2
\ee
This instanton yields the amplitude for pair production in the presence of an electric field. Their action provides the dominant term in the quantitative formula for this amplitude.

\begin{center}
\includegraphics[width=0.49\linewidth]{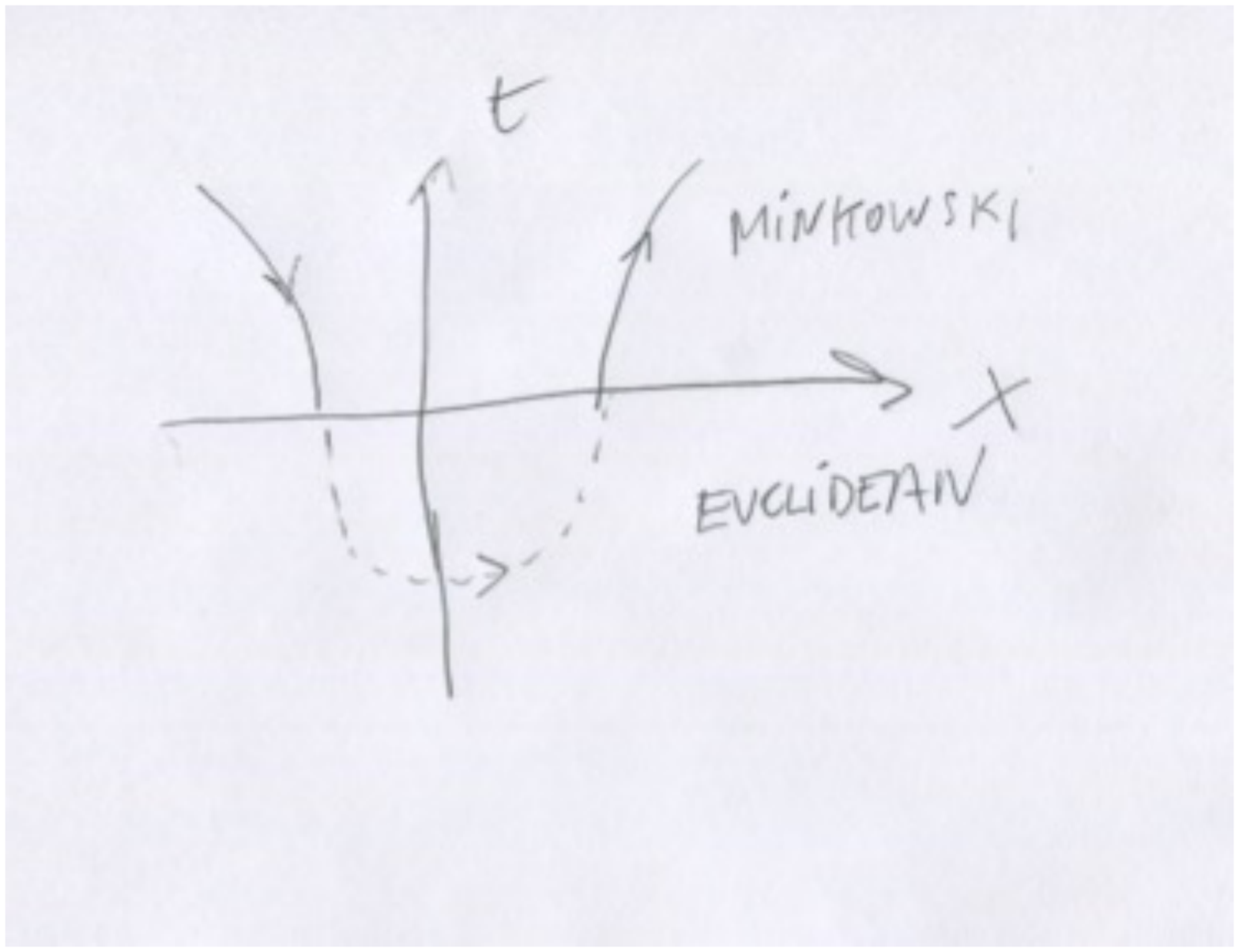}
\captionof{figure}{Pair creation by an electric field.}
\end{center}

Proceeding by analogy, Alex Vilenkin tried to  apply this idea in the creation of universes from nothing. Assume the familiar Friedmann metric for space-time
\be
ds^2= dt^2-a^2(t) \d_{ij} dx^i dx^j
\ee

The lorentzian solution for the scale factor of de Sitter space of radius $L={1\over H}$ is give ny
\be
a(t)={1\over H}\cosh\, Ht
\ee
where $H^2$ is proportional to the vacuum energy density. The corresponding euclidean solution (the de Sitter instanton of Gibbons and Hawking \cite{GH}) is
\be
a(\t)={1\over H}\cos\, H\t
\ee
Vilenkin interpreted this instanton by analogy with the Schwinger process by indicating an amplitude for the creation of the universe from nothing whatsoever. The euclidean manifold is glued to the lorentzian one at a {\em moment of time symmetry} where the extrinsic curvature vanishes.

\begin{center}
\includegraphics[width=0.49\linewidth]{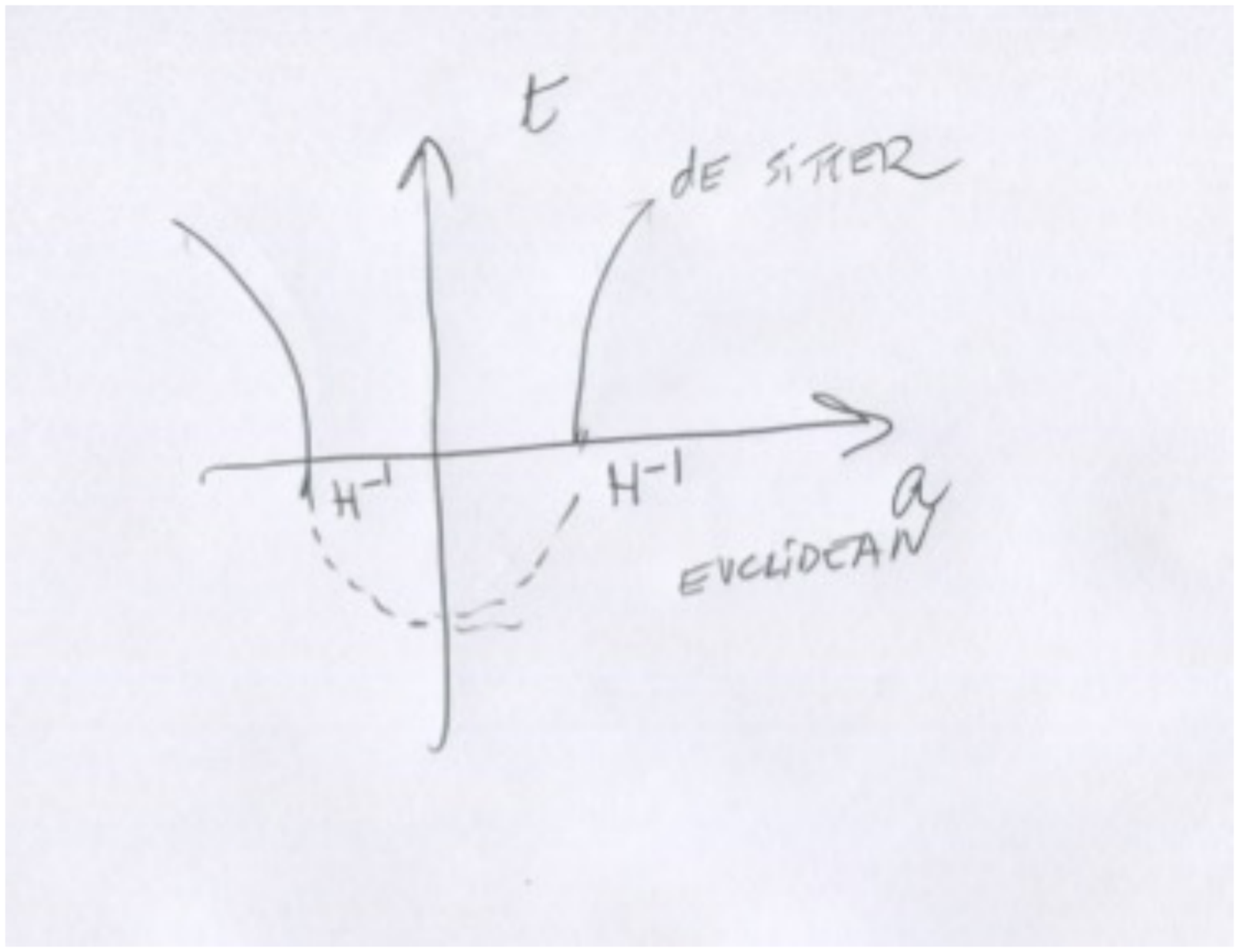}
\captionof{figure}{Creation of a universe out of nothing.}
\end{center}

We have already pointed out the difficulties of physical interpretation of the wave function $\psi[h_{ij}]$ in quantum gravity. Nevertheless Hartle and Hawking proposed a concrete way to compute this {\em wavefunction of the universe}. The no-boundary state (now best known as the Hartle-Hawking state) is characterized by the wave functional (that yields  some amplitude for a three-manifold $\Sigma$ endowed with  the metric $h_{ij}$), and some corresponding values of the matter fields, $\phi$ computed by the functional integral
\be
\psi[ h_{ij},\phi]\equiv \int {\cal D} g_{\m\n}\, {\cal D}\phi \,e^{-S}
\ee
where $\phi$ represents all matter fields and the functional integral is made over all four-dimensional manifolds whose unique boundary is $\Sigma$ characterized by the matrix $h_{ij}$. This is the origin of the name {\em no boundary state}. It has no more boundaries than absolutely necessary. 
\par
It so happens that every three-dimensional closed surface is null-cobordant, which means that it is the boundary of a four-dimensional manifold.
In general cobordism classes are given by the Stiefel-Whitney characteristic numbers.
\par
When expressing the functional integral using the ADM decomposition, the fact that the lapse $N$ is a gauge artifact means that
\be
\int {\cal D} g_{\m\n}\, {\cal D}\, \phi \,{d\over d N}\,e^{-S}=0
\ee
It can be easily shown that this condition formally reproduces the Wheeler-DeWitt equation for the wave function of the universe. This one of the main virtues of the Hartle-Hawking state.

\section{Symmetries and observables}
It is generally accepted that General Relativity, a generally covariant theory, is akin to a gauge theory, in the sense that the diffeomorphism group of the apace-time manifold, $Diff(M)$ plays a role similar to the compact gauge group in the standard model of particle physics. This symmetry, or invariance under coordinate changes, is a reflection of the fact that GR is a geometric theory; all concepts and equations can be written in a geometric way. There are some differences though.
\par
 To begin with, the group, $ Diff(M)$ is too large; is not even a Lie group \cite{Milnor}. Besides, its detailed structure depends on the manifold, which is a dynamical object not given a priori: it has to be determined by Einstein's equations, once given the material content. Other distinguished subgroups (such as the area-preserving diffeomorphisms \cite{Alvarezz}) are perhaps also arguable for. These diffeomorphims are such that
 \be
 \pd_\l \xi^\l=0
 \ee
 such that they are sometimes referred to as {\em transverse diffeomorphisms}. In a Taylor expansion in a local patch
 \be
 \xi^\m\equiv \,_0 \xi^\m+\,_0\xi^\m_\a\, x^\a+\,_0\xi^\m_{\a\b}\, x^\a x^\b+\ldots
 \ee
 (where $\,_0\xi^\m_{\a_1\a_2\ldots\a_p}$ are arbitrary constants)  this means that all coefficients are totally traceless.
 \be
 \,_0\xi^\m_\m=\,_0\xi^\m_{(\m\b)}=\ldots=0
 \ee
 That is
 \bea
 &\,_0\xi^\m_\a\rightarrow\,_0\xi^\m_\a-{1\over n}\xi^\l_\l\, \d^\m_\a\nonumber\\
 &\,_0\xi^\m_{\a\b}\rightarrow\,_0\xi^\m_{\a\b}-{1\over n+1}\left(\,_0\xi^\l_{\l\b}\,\d^\m_\a+\,_0\xi^\l_{\a}\,\d^\m_\b\right)
 \eea
 and so on. Einstein himself was fond of  the gauge condition $|g|=1$; and was the first to propose some incomplete version of the unimodular theory \cite{E19}\cite{unimodular}. Those area preserving diffeomorphisms leave invariant a given measure, such as the Lebesgue measure, $d^n x$, although they share  problems with their bigger cousin $Diff(M)$.
\par
Fock proposed a long time ago \cite{Fock} that {\em harmonic coordinates} should be privileged, and that in some sense they were the only with a physical meaning. They are defined by
\be
\Box x^{(\l)}=0
\ee
(where the coordinates ere considered as a set of functions ob the manifold
\be
x^{(\l)}: M\rightarrow \mathbb{R}^n
\ee
At the linear level this is equivalent to the de Donder gauge condition,
\be
\pd_\m h^{\m\l}={1\over 2}\pd^\l h
\ee
This condition is left invariant by linearized diffeomorphims such that the generatic vector is harmonic
\be
\Box \xi^\l=0
\ee
that is; all contractions of the flat metric with the covariant indices should vanish
\be
\eta^{\a\b}\,_0\xi^\m_{\a\b}=\eta^{\a\b}\,_0\xi^\m_{(\a\b\g)}\ldots=0
\ee
That is
\bea
&\,_0\xi^\m_{\a\b}\rightarrow \,_0\xi^\m_{\a\b}-{1\over n}\eta_{\a\b}\, \eta^{\r\s}\,_0\xi^\m_{\r\s}\nonumber\\
&\,_0\xi^\m_{(\a\b\g)}\rightarrow\,_0\xi^\m_{(\a\b\g)}-{1\over n+2}\left(\eta_{\a\b} V^\m_\g+\eta_{\a\g} V^\m_\b+\eta_{\b\g} V^\m_\a\right)
\eea
(where $V^\m_\a\equiv \eta^{\r\s}\, \,_0\xi^\m_{\r\s\a}$) and so on.
\par
It also seems clear that when we are integrating upon a restricted class of spacetimes with some specific type of boundary, or asymptotic behavior, then the gauge group is restricted to the subgroup consisting on those diffeomorphisms that act trivially on the boundary (or leave invariant the boundary conditions). The subgroup that act not-trivially is related to the set of conserved charges, if any. In the asymptotically flat case this is precisely the Poincar\'e group, $SO(1,3)$ that gives rise to the ADM mass and also the BMS group acting on null infinity.
\par
In the asymptotically anti-de Sitter case, this is related to the conformal group $SO(2,3)$. 
\par
It is nevertheless not clear what is the physical meaning of keeping constant the boundary of spacetime (or keeping constant some set of boundary conditions) in a functional integral of some sort. This is related to the issue of whether the functional integral over geometries allows for topology change.
\par
Incidentally, it is very difficult to define what could be {\em observables} in a diffeomorphism invariant theory, other than global ones defined as integrals of scalar densities composite operators $O(\phi_a(x))$ (where $\phi_a, a=1\ldots N$ parametrizes all physical fields) with the peudo-riemannian measure
\[
{\cal O}\equiv \int \sqrt{|g|}\,d^4 x\, O(\phi_a(x))
\]
 Some people claim that there are no local observables whatsoever, but only {\em pseudolocal } ones \cite{Giddings}; the fact is that we do not know.
 Again, the exception to this stems from keeping the boundary conditions fixed; in this case it is possible to define an $S$-matrix in the asymptotically flat case, and a conformal quantum field theory (CFT) in the asymptotically anti-de Sitter case. 
 Unfortunately, the most interesting case from the cosmological point of view, which is when the space-time is asymptotically de Sitter is not well understood.
\par
In the mathematical front, it is well known that the equivalence problem in four-dimensional geometries is undecidable \cite{Alvarez}. This theorem, which was first proved by Kolmogorov, states that given two four-dimensional manifolds, there is no systematic procedure to determine whether those two manifolds are diffeomorphic or not. In three dimensions  Thurston's geometrization conjecture has recently been put on a firmer basis by Hamilton and Perelman, but it is still  not clear whether it can be somehow implemented in a functional integral without some drastic restrictions. Those caveats should be kept in mind when reading the sequel. 
\par
Gauge theories can be formulated in the {\em bakground field approach}, as
introduced by B. DeWitt and others (cf. \cite{DeWitt}). In this approach,
the quantum field theory depends on a background field, but not on any one in particular,
and the theory enjoys background gauge invariance.
\par
Is it enough to have the functional integral of quantum gravity formulated in such a way? 
\par
It can be argued that the only 
vacuum expectation value consistent with diffeomorphism invariance is
\be
\langle 0\left|g_{\a\b}(x)\right|0\rangle=0
\ee
in which case the answer to the above question ought to be in the negative, because 
this is a singular
background and curvature invariants do not make sense.
It all boils down as to whether the ground state of the theory is diffeomorphism
 invariant or not.
There is an example, namely three-dimensional gravity in which invariant quantization 
can be performed
\cite{Witten3}. In this case at least, the ensuing theory is almost topological, although the issue is not completely clear owing to subleties related to the Ba\~nados-Henneaux-Teitelboim (BHT) blach hole. 
\par
In all attempts of a canonical quantization of the gravitational field, 
one always ends up with an (constraint) equation
corresponding physically to the fact that the total hamiltonian of a parametrization 
invariant theory
should vanish. When expressed in the Schr\"odinger picture, this equation is often 
dubbed the {\em Wheeler-de Witt equation}. This equation is plagued by
operator ordering and all other sorts of ambiguities.
It is curious to notice that in ordinary quantum field theory there also exists a Schr\"odinger
representation, which came recently to be controlled well enough as to be able to 
perform lattice 
computations (\cite{Makeenko}).
\par
Gauge theories can be expressed in terms of gauge invariant operators, such as
Wilson loops . They obey a complicated
set of equations, the loop equations, which close in the large $N$ limit
as has been shown by Makeenko and Migdal (\cite{Makeenko}). These equations can be properly 
regularized,
e.g. in the lattice. Their explicit solution
is one of the outstanding challenges in theoretical physics. Although many conjectures
have been advanced in this direction, no definitive result is available.

\section{Do Strings answer any of our questions?}
It should be clear by now that we probably still do not know
what is exactly the problem to which string theories \cite{Green} \cite{Polchinskis} are the answer. This fact has been repeatedly emphasized by the late Joe Polchinski \cite{Polchinski}. We shall concentrate in just one aspect that we believe to be important and which turns out to be quite contentious. Ever since Maldacena's conjecture (more on this in a moment) some people put forward the idea the gravity is emergent in a holographic way from a conformal field theory (CFT) defined in the boundary of the bulk spacetime (one dimension less).
This is a fascinating topic, which drives much of the research in the field. We shall give here just the general idea, and then comment on some aspects of it. Perhaps just one comment on the way quantum gravity appears in string theory.
\par
 The starting point of the whole topic is a one-dimensional object living in some D-dimensional flat space. Consistency of the quantization is believed to be possible only when conformal symmetry is maintained in the process. This implies that $D=26$. Absence of tachyons is only possible (through a projection first invented by Gliozzi, Scherk and Olive (GSO)) \cite{Gliozzi} when supersymmetry is implemented and this implies $D=10$.
 \par
  There are two places in which gravitation enters into the game.
  \par
 There are two types of strings: closed and open. Open strings also include closed strings as a subsector; but the opposite is not true: there are consistent theories of closed strings only. When the spectrum of states is analyzed, one finds gauge fields in the open string sector, and gravitons in the closed string sector. One can indeed study on-shell perturbative string amplitudes, and take the limit of infinite string tension (where strings degenerate into points) and in this limit one gets quantum field theory (QFT) amplitudes (usually with much supersymmetry).
 \par
 There is another way to understand gravitation in the framework of string theory. We can try to understand the quantum behavior of strings in some curved space-time. Demanding conformal invariance (the vanishing of the corresponding beta functionals) \cite{Friedan} imply that the background has to obey some field equations, which correspond to some (supersymmetric) generalization of Einstein's equations.
 
 Finally, there is a still more indirect way. As we shall see in a moment, there are indications of some dualities strong/weak voupling in string theories. They suggest that there is some as yet unknown theory (dubbed {\em M-theory}) which explains all these symmetries. Not much is known about this theory, unfortunately . One of the main problems is that it is always strongly coupled; there is no weak coupling regime. Suggestive as all this might be, we lack ideas on how to do convincing computations. 
 \par
 Even less is known of how to make contact with the low-energy, non-supersymmetric world. Here the main problem is the huge arbitrariness that necessarily enters when breaking supersymmetry. This implies a huge loss of predictivity.

\subsection{The Maldacena conjecture and gravity/CFT duality.}
 Maldacena \cite{Maldacena} proposed as a conjecture that $IIB$ string theories in a background
$AdS_5\times S_5$ with  common radius $L\sim l_s (g_s N)^{1/4}$ (where $l_s$ is the characteristic length of string theory defined by the string tension through $\a^\prime\equiv {1\over l_s^2}$, and $g_s$ is the string coupling constant, related to the value of the dilaton field)  and N units of RR flux 
that is,  $\int_{S_5} F_5 =N$ (which implies that $F_5\sim \frac{N}{r^5}$)
is equivalent to a four dimensional ordinary gauge theory in flat four-dimensional Minkowski
space, namely ${\cal N}=4$ super Yang-Mills with gauge group $SU(N)$ and coupling constant
$g=g_s^{1/2}$.
 \par

Although there is much supersymmetry in the problem and the kinematics largely determine 
correlators,
(in particular, the symmetry group $SO(2,4)\times SO(6)$ is realized as an isometry group on the
gravity side and as an $R$-symmetry group as well as conformal invariance on the gauge 
theory side)
this is not fully so \footnote{The only correlators that are completely determined
through symmetry are the two and three-point functions.}
and the conjecture has passed many tests in the semiclassical approximation
to string theory, which corresponds to large ${L\over l_s}$, dual to large $N$ on the CFT side.

The action of the RR field, given schematically by $\int F_5^2$, scales as $N^2$, whereas the 
ten-dimensional 
Einstein-Hilbert $\int R$, depends on the overall geometric scale as the eighth power of the
common radius, $L^8$. The 't Hooft coupling is
 $\lambda = g^2 N \sim \frac{L^4}{l_s^4}$  and the tenth dimensional Newton's constant
is  
\be
\kappa_{10}^2\sim G_{10}\sim l_p^8=g_s^2 l_s^8\sim\frac{L^8}{N^2}.
\ee
\par  
 If we consider the effective five dimensional theory after compactifying 
on a five sphere of radius $r$, the RR term yields a negative contribution $\sim -
(\frac{N}{r^5})^2 r^5$ ,whereas the positive curvature of the five sphere $S^5$ gives a positive
contribution, $\sim \frac{1}{r^2} r^5$. The competition between these two terms in the effective
potential is responsible for the minimum with negative cosmological constant.

\par
The way the dictionary works  in detail \cite{Witten} is that the supergravity action 
corresponding to fields 
with prescribed
boundary values is related to gauge theory correlators of certain gauge invariant operators 
corresponding to the particular field studied:
\be
\left.e^{- S_{sugra}[\Phi_i]}\right|_{\left|\Phi_i\right|_{\partial AdS}=\phi_i}= 
\left\langle e^{\int {\cal O}_i\phi_i}\right\rangle_{CFT}
\ee

\par
 This is the first time that a precise holographic description of spacetime in terms of a 
(boundary) 
gauge theory is proposed and, as such it is of enormous potential interest. 
It has been conjectured 
by 't Hooft \cite{'tHooft} and further developed by Susskind \cite{Susskind} that there 
should be
much fewer degrees of freedom in quantum gravity than previously thought. The conjecture 
claims that
it should be enough with one degree of freedom per unit Planck surface in the two-dimensional
 boundary
of the three-dimensional volume under study. The reason for that stems from an analysis of the 
Bekenstein-Hawking \cite{Bekenstein}\cite{Hawking} 
entropy associated to a black hole, given in terms of the
two-dimensional area $A$ \footnote{The area of the horizon for a Schwarzschild black hole 
is given by:
\be
A=\frac{8\pi G^2}{c^4}M^2
\ee
}
of the horizon by
\be
S=\frac{ c^3}{4 G\hbar}A.
\ee

This is  a deep result indeed, still not fully understood, although in the particular case of extremal black holes (the only ones that are compatible with supersymmetry) the dependence of the entropy with the full set of charges has been reproduced by a remarkable string theory calculation by Strominger and Vafa \cite{Strominger}.
\par
It is true on the other hand that the Maldacena conjecture has only been checked for the 
time being 
in some corners of 
parameter space corresponding to the semiclassical approximation, namely when strings can be approximated by supergravity in the appropiate 
background.

\subsection{String Dualities and branes}
The so- called {\em T-duality} is the simplest of all dualities and the 
only one which can be shown
to be true, at least in some contexts\cite{PolchinskiD}. At the same time it is a very stringy
characteristic, and depends in an essential way on strings being extended
objects.  In a sense, the web of dualities
rests on this foundation, so that it is important to understand 
clearly the basic 
physics involved. Let us consider strings living on an external space with one 
compact dimension, which we shall call $Y$, with topology $S^1$ and radius $R$. 
The corresponding field in the imbedding of the closed
string (where we identify in the word sheet of the string the spatial coordinate $\s\sim \s+2\pi$), which we shall still  call $Y$
({\it i.e.} we are dividing the target-spacetime dimensions as $(X^{\m},Y)$,
where $Y$ parametrizes the circle), has then the
possibility of winding around it:
\be
Y(\sigma + 2\pi,\tau) = Y(\sigma, \tau) + 2\pi R m \; .
\ee
A closed string can close in general up to an isometry of
the external spacetime.
\par
The zero mode expansion of this coordinate (that is, forgetting about
oscillators) would then be
\be
Y = Y_{c} + 2 P_{c} \tau + m R \sigma \; .
\ee
Canonical quantization leads to $[Y_{c},P_{c}] = i$, and single-valuedness
of the plane wave $e^{i\, Y_{c}P_{c}}$ enforces as usual $P_{c}\in \mathbb{Z}/R$,
so that $P_{c} = \frac{n}{R}$.
\par
The zero mode expansion can then be organized into left and right movers in the
following way
\bea
Y_L (\tau +\sigma)&=& Y_c/2 + 
     \left(\frac{n}{R} + \frac{mR}{2}\right) (\tau + \sigma) \; ,\nonumber\\
Y_R (\tau -\sigma)&=& Y_c/2 + 
    \left(\frac{n}{R} - \frac{mR}{2}\right) (\tau - \sigma) \; .
\eea
The mass shell conditions reduce to
\bea
m_L^2 &=& \frac{1}{2}\left(\frac{n}{R} + 
          \frac{mR}{2}\right)^2 + N_L -1 \; ,\nonumber\\
m_R^2 &=& \frac{1}{2}\left(\frac{n}{R} - \frac{mR}{2}\right)^2 + N_R -1 \; .
\eea
Level matching, $m_L = m_R$, implies that there is a relationship between
momentum and winding numbers on the one hand, and the oscillator 
excess on the other
\be
N_R - N_L \;=\;  nm \; .
\ee
At this point it is already evident that the mass formula is invariant
under
\be
R\; \rightarrow\;  R^{*}\equiv {2 l_s^2\over R} \; ,
\ee
provided that at the same time one exchanges momentum and winding numbers. This is the simplest instance of 
{\em T-Duality}.
\par
On the other hand, it is an old observation (which apparently originated in 
Schr\"odinger) that Maxwell's 
equations are almost symmetrical with respect to
interchange between electric and magnetic degrees of freedom (electromagnetic duality). This idea was explored by Dirac
and eventually lead to the discovery of the consistency conditions between electric and magnetic charges that have to be fulfilled
if there are magnetic monopoles in nature. The fact that nonsingular magnetic monopoles 
appear as classical solutions in some gauge theories led further support to this duality
viewpoint. In order to be able to make a consisting conjecture, first put forward by Montonen
and Olive \cite{Montonen}, supersymmetry is needed, as first remarked by Osborn \cite{Osborn}. 
\par
Now in strings there are the so- called Ramond-Ramond (RR) fields, which are p- forms
of different degrees. In the same way that one forms (i.e., the Maxwell field) couples to 
charged particles  that is, from the spacetime point of view, to objects of dimension 0
with one-dimensional trajectories, a p-form
\be
A_{\m_1\ldots\m_p}
\ee
would couple to a $(p-1)$-dimensional object, whose world history is described by a 
$p$-dimensional
hypersurface
\be
x^{\m}= x^{\m}(\xi_1\ldots \xi_p)
\ee
These objects are traditionally denoted by the name $p$-branes (it all originated in a 
dubious joke). That is, ordinary particles are $0$-branes, a string is a $1$-brane, a membrane
is a $2$-brane, and so on.
\par
Dualities relate branes of different dimensions in different theories; this means that
if one is to take this symmetry seriously, it is not clear at all that strings are the
more fundamental objects: in the so called $M$-theory (to be introduced in a moment) branes appear as fundamental as strings.
\par
If we are willing to make the hypothesis
that supersymmetry is not going to
be broken whilst increasing the coupling constant, $g_s$,
some astonishing conclusions
can be drawn. Assuming this, massless quanta can become massive
as $g_s$ grows only if their number, charges and spins are such that
they can combine into massive multiplets  
(which are all larger than the irreducible massless ones). 
The only remaining issue, then, is whether
any other massless quanta can appear
at strong coupling.
\par
Now, in the IIA string theory there are states associated to the Ramond-Ramond (RR)
one form, $A_1$, namely the D-0-branes, whose tension goes as 
$m\sim\frac{1}{g_s}$. This clearly gives new massless states in the strong 
coupling limit.
\par
There are reasons\footnote{In particular: The fact
that there is the possibility of a central extension in the IIA algebra,
related to the Kaluza-Klein compactification of the d=11 Supergravity
algebra.} to think that this new massless states  are the first
level of a Kaluza-Klein tower associated to compactification on a circle
of an 11-dimensional theory.
Actually, assuming an 11-dimensional spacetime with an isometry 
$k=\frac{\partial}{\partial y}$, an Ansatz which exactly reproduces the 
dilaton factors of the IIA string is
\be
ds^{2}_{(11)} = e^{\frac{4}{3}\phi}(dy - A^{(1)}_{\m} dx^{\m})^2 + 
e^{-\frac{2}{3}\phi}g_{\m\n}dx^{\m}dx^{\n} \; .
\ee 
Equating the two expressions for the D0 mass,
\be
\frac{1}{g_s} = \frac{1}{R_{11}} \; ,   
\ee
leads to $R_{11} = e^{\frac{2}{3}\phi} = g_A^{2/3}$ (where $g_A$ is the gauge coupling constant).
\par
This means that a new dimension appears at strong coupling, and this dimension is 
related to the dilaton. The only reason why we do not see it at low energies is 
precisely because of the smallness of the string coupling, related directly to the
dilaton field.
The other side of this is that this eleven dimensional theory, dubbed {\em M-theory}
does not have any weak coupling limit; it is always strongly coupled. Consequently, not much
is known on this theory, except for the fact that its field theory, low curvature limit
is ${\cal N}=1$ supergravity in $d=11$ dimensions.
\par
All supermultiplets of massive one-particle states of the IIB string
supersymmetry algebra contains states of at least spin 4. 
This means that under the previous set of hypothesis, the set of 
massless states at weak coupling must
be exactly the same as the corresponding set at strong coupling.
This means that there must be a symmetry mapping weak coupling into 
strong coupling.
\par
There is a well-known candidate for this symmetry:
Let us call, as usual, $l$ the RR scalar and $\phi$ the dilaton (NSNS).
We can pack them together into complex scalar
\be
S \;=\;  l \,+\, i e^{-\frac{\phi}{2}} \; .
\ee
The IIB supergravity action in d=10 is invariant under the $SL(2,\mathbb{R})$
transformations
\be
S \rightarrow \frac{a S + b}{ c S + d} \; ,
\ee
if at the same time the two two-forms, $B_{\m\n}$ (the usual, ever-present,
NS field), and $A^{(2)}$, the RR field transform as
\bea
\left( \begin{array}{c}
B\\
A^{(2)}
\end{array}\right ) \rightarrow 
\left( \begin{array}{cc}
d & - c \\
- b & a\end{array}\right )
\left ( \begin{array}{c}
B\\
A^{(2)}
\end{array}\right) \; , 
\eea
Both the, Einstein frame, metric $g_{\m\n}$
and the four-form $A^{(4)}$ are inert under this
$SL(2,\mathbb{R})$ transformation.
\par
A discrete subgroup $SL(2,\mathbb{Z})$ of the full classical $SL(2,\mathbb{R})$
is believed to be an exact symmetry of the full string theory. 
The exact imbedding of the discrete subgroup in the full $SL(2,\mathbb{R})$ 
depends on the
vacuum expectation value of the RR scalar.
\par
The particular transformation
\bea
g = \left(
\begin{array}{cc}
 0& 1\\
- 1& 0\end{array}
\right) \; ,
\eea
maps $\phi$ into $ - \phi$ (when $l =0$),
and $B$ into $A^{(2)}$. 
This means that the string coupling
\be
g_s\rightarrow \frac{1}{g_s}
\ee
This is a strong/weak coupling type of duality, similar to the electromagnetic duality
in that sense .The standard name for it is  an 
{\em S-duality} type of transformation, mapping the ordinary string with NS charge,
to another string with RR charge (which then must be a $(D-1)$-brane,
and is correspondingly called a {\em D-string}), and, from there, 
is connected to all other D-branes by T-duality.
\par

Using the fact that upon compactification on $S^1$, IIA at $R_A$ is
equivalent to IIB at $R_B\equiv 1/R_A$, and the fact that the effective action
carries a factor of $e^{-2\phi}$ we get
\be
R_A g_B^2 = R_B g_A^2 \; ,
\ee
which combined with our previous result, $g_A = R_{11}^{3/2}$ implies 
that $ g_B = \frac{R_{11}^{3/2}}{R_A}$.
Now the Kaluza-Klein Ansatz implies that from the eleven dimensional 
viewpoint the compactification radius is measured as
\be
R_{10}^2 \equiv R_A^2 e^{- 2\phi/3} \; ,
\ee
yielding
\be
g_B = \frac{R_{11}}{R_{10}} \; .
\ee
\par
 From the effective actions written above it is easy to check 
that there is a (S-duality type) field transformation mapping
the SO(32) Type I open string into the SO(32)
Heterotic one namely
\bea
g_{\m\n}&&\rightarrow e^{-\phi} g_{\m\n}^{Het} \; ,\nonumber\\
\phi&&\rightarrow - \phi \; ,\nonumber\\
B'&&\rightarrow B \; .
\eea
This means that physically there is a strong/weak coupling duality, because 
coupling constants of the compactified theories  would
be related by
\bea
g_{het}&& = 1/g_I \; ,\nonumber\\
R_{het}&& = R_I/g_I^{1/2} \; .
\eea
\par
\begin{center}
\includegraphics[width=0.49\linewidth]{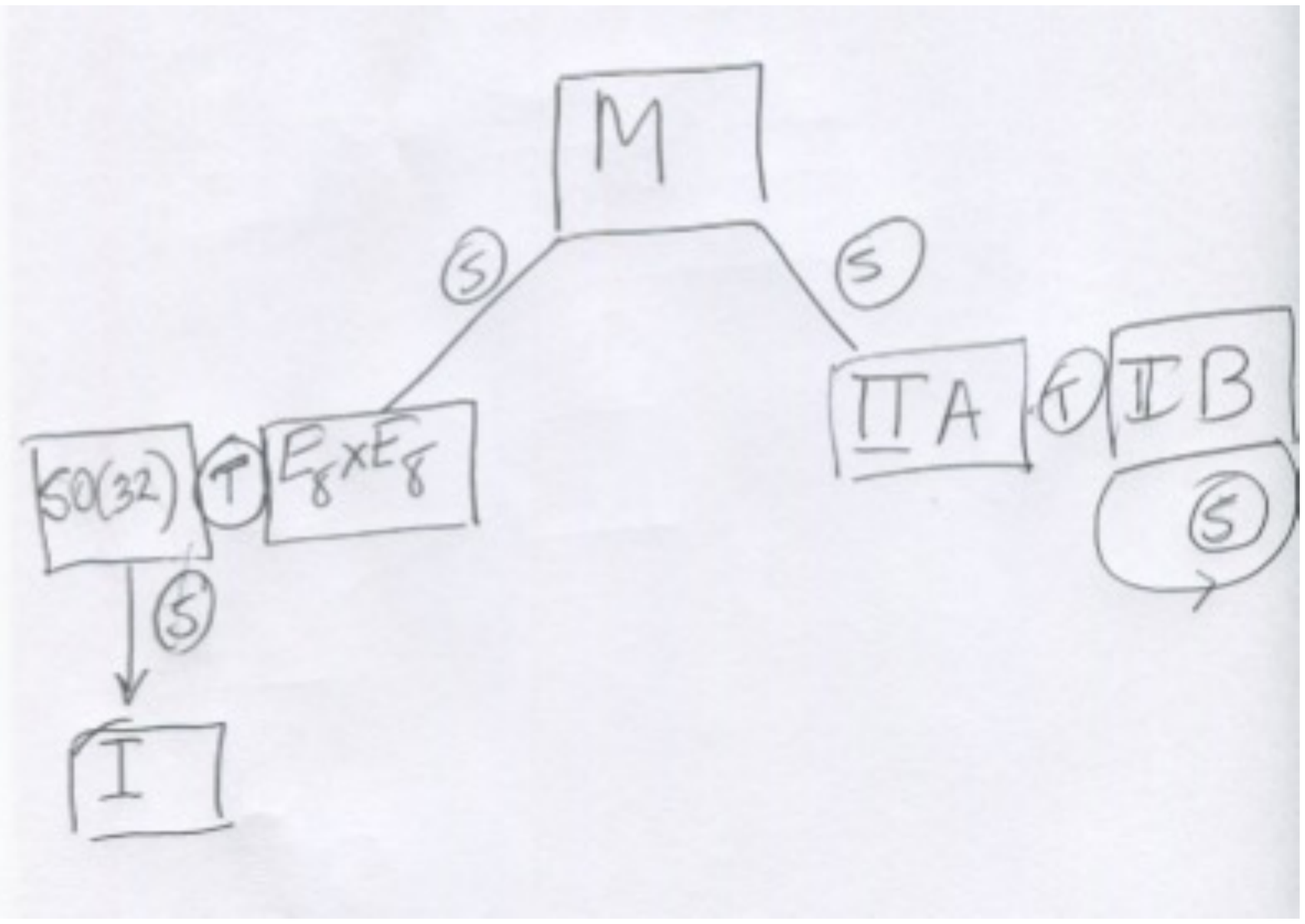}
\captionof{figure}{All string theories are related by T or S-dualities.}
\end{center}

We have only scratched the surface here, but we can refer to the several excellent books treating this subject in more depth such as for example \cite{Tomas}.

\subsection{It from bit. Spacetime= entanglement, ER=EPR, and all that.}
Ever since Wheeler \cite{Wheeler} wrote the sentence "...every physical quantity, every it derives its ultimate significance from bits, binary yes-or no indications, a conclusion which we epitomize in the phrase {\em it from bit} ", many people have attempted to give a precise meaning to this idea.
\par
\begin{center}
\includegraphics[width=0.49\linewidth]{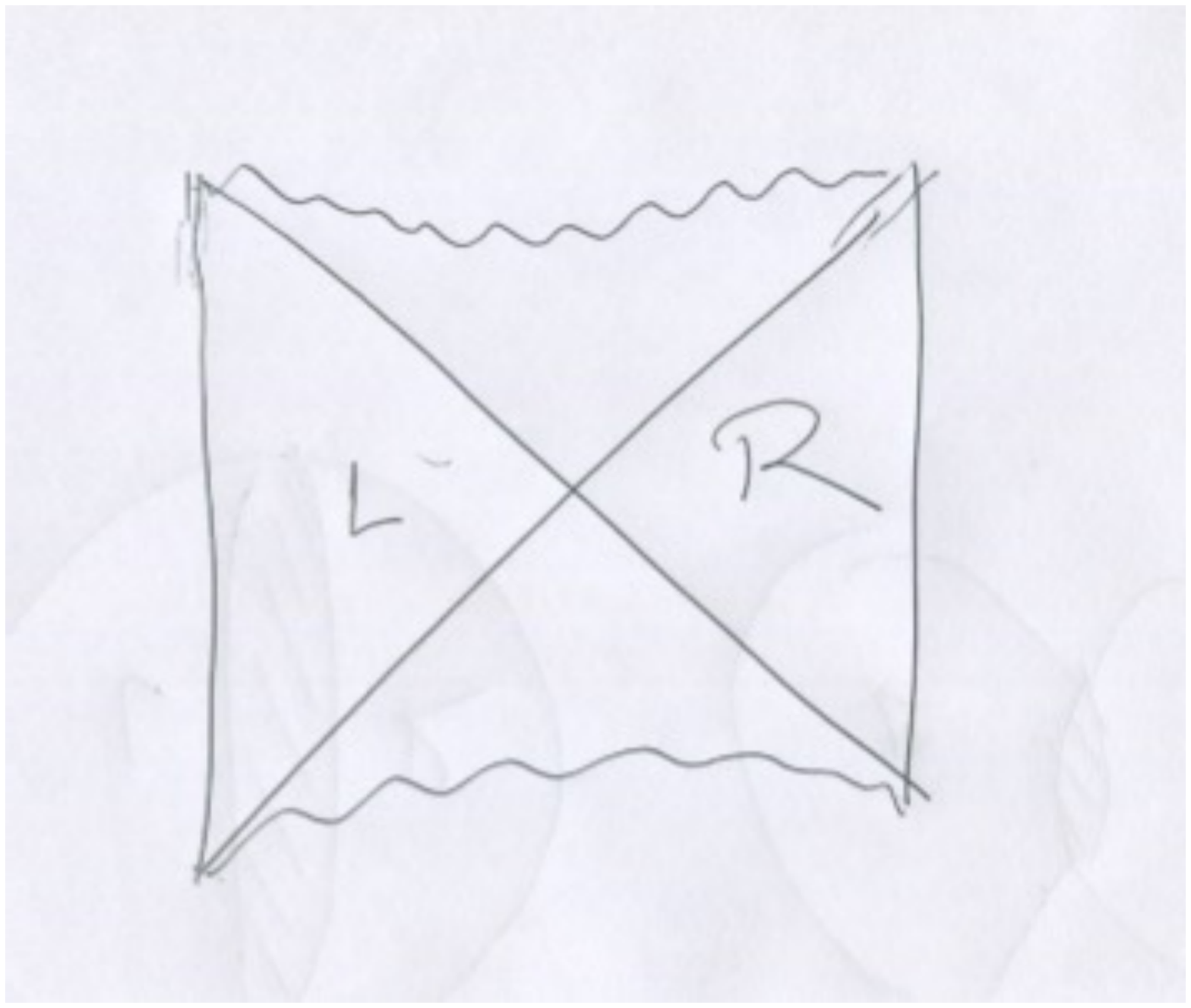}
\captionof{figure}{Eternal AdS black hole.}
\end{center}

For example, van Raamsdonk \cite{VanRaamsdonk} has proposed a radical reinterpretation in the context of gauge/gravity duality of Israel's \cite{Israel} thermofield black hole state
\be
|\psi_\b\rangle\equiv \sum_n e^{-\b{E_n\over 2}}|E_n\rangle_L \otimes |E_n\rangle_R
\ee
as representing the eternal AdS-Schwarzschild black hole. 
As is well-known the {\em thermofield state} \cite{Takahashi} is a clever form of reinterpreting statistical mechanics by postulating another copy of the physical Hilbert space, and consider the product Hilbert space
\be
H_{extended}\equiv H_{phys}\otimes H_{copy}
\ee
In this extended Hilbert space we consider the thermofield state
\be
|\psi_\b\rangle\equiv \sum_{n} e^{-\b{E_n\over 2}}|E_n^{phys}\rangle\otimes |E_n^{copy}\rangle
\ee
It so happens that when computing the expectation value of a physical observable (which acts solely oh $H_{phys}$) on the extended Hilbert space, we get
\bea
&\left\langle \psi_\b\left|{\cal O}^{phys}\right|\psi_\b\right\rangle= \sum_{n,m}\left(e^{-\b{E_n\over 2}}\langle E_n^{phys}|\otimes \langle E_n^{copy}|\right){\cal O}^{phys}\left(| e^{-\b{E_m\over 2}}|E_m^{phys}\rangle\otimes |E_m^{copy}\rangle\right)=\nonumber\\
&=\sum_n e^{-\b E_n}\langle E_n|{\cal O}^{phys}|E_n\rangle\equiv \text{tr}\,\left(\r_\b {\cal O}^{phys}\right)
\eea
owing to the assumed orhogonality of the set of states
\be
\langle E_n|E_m\rangle=\d_{nm}
\ee
and $\r_\b\equiv e^{- \b H_{phys}}$ is the thermal density matrix.
\par
The rationale of the reinterpretation is as follows. Assume that the thermofield state is in some sense a true state with a gravity dual.  The gravity dual should be a spacetime with two asymptotically AdS regions, each of them corresponding to a copy of the CFT. In one of this asymptotic regions the dual should be just an AdS black hole, which is just a thermal state of the CFT. This is consistent with the fact that, as we have just seen,
\be
\text{tr}_L\,|\psi_\b\rangle\langle\psi_\b|=\sum_n e^{-\b E_n} |E_n\rangle_R\langle E_n|_R\equiv\r_\b
\ee
The presence of horizons in the gravity dual that prevent classical communication between both asymptotic regions are consistent with the absence of interactions between both copies of the CFT.
\par
This means that the state $|\psi_\b\rangle$ which could be naively though as being dual to a superposition of disconnected spacetimes, is reinterpreted as a classically connected spacetime.
In this example, connectivity stems from entanglement of the two components present in the thermofield state.
\par
Another argument starts by considering a CFT on a sphere, which we again divide in two different hemispheres, which we keep denoting by $L$ and $R$. Assume that the physical Hilbert space is of the form
\be
H=H_L\otimes H_R
\ee
(more on this later).
\begin{center}
\includegraphics[width=0.49\linewidth]{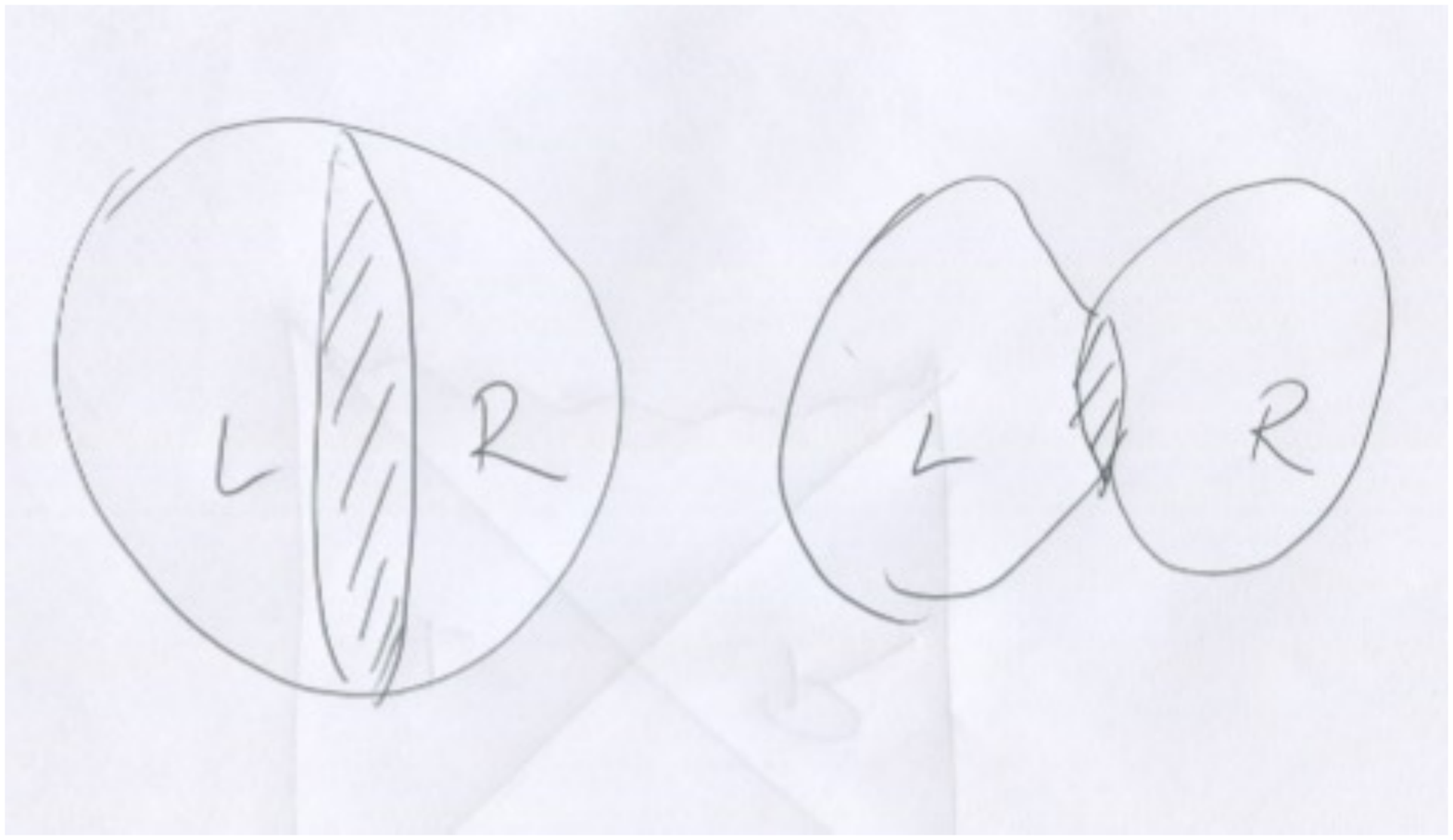}
\captionof{figure}{Sphere divided intotwo hemispheres.}
\end{center}

Assume also that a measure of entanglement between both hemispheres is given by the entanglement (von Neumann) entropy associated to the physical state $|\Psi\rangle$
\be
S(L)\equiv -\text{tr}\,\r_L\,\log\,\r_L
\ee
where
\be
\r_L\equiv \text{tr}_R\,|\Psi\rangle\langle\Psi|
\ee 
This quantity should be regularized. In fact, in the context of AdS/CFT duality Ryu and Takayanagi \cite{Ryu} have proposed that the entropy associated to a finite area $A$ in the CFT side (defined on the boundary $\pd M$ of the bulk  space-time $M$) is given by area of the minimal surface $S$ in the bulk $M$ whose boundary on $\pd M$ is precisely $A$: $\pd S=A$, the area under consideration in the CFT.
\par
 Assume that this entropy $S(L)$ decreases. Then according to Ryu and Takayanagi's formula, the area of the minimal surface that separates both components $L$ and $R$ should decrease as well. This means that as the entanglement goes to zero, both regions $L$ and $R$ are pinching off, and in the limit we get two disconnected pieces.
\par
This viewpoint has been forcefully supported by Maldacena and Susskind \cite{Maldacena}  that proposed that the gravity dual of quantum mechanical entanglement (symbolically, EPR, for Einstein, Podolsky and Rosen \cite{EPR}) should involve wormholes (symbolically, ER, for Einstein-Rosen bridges \cite{ER}). This is the famous slogan ER=EPR.  In this way they are able to question the necessity of firewalls \cite{firewalls} in some cases by considering the entanglement between the black hole and the radiation after the Page time.
\par
General attempts of this type to define spacetime starting from entanglement have been criticized \cite{Giddingss} on the basis  that in order to define entanglement one needs to define a notion of subsystems and entanglement entropy depends on such definition of subsystems.
\par
Some people still maintains that spacetime structure appears more fundamental than entanglement.
\subsection{Diffeomorphism invariant observables.}
Dirac himself \cite{Diracc} worried about the general problem of how to define gauge invariant observables in the context of (abelian) gauge theories and introduced {\em dressed operators} such as, for a charged scalar field,
\be
\Phi_D(x)\equiv\phi(t,\vx)\,e^{i C(x)}
\ee
He assumed that
\be
C(x)\equiv \int c_i(x,\xp) A^i(\xp) d\xp
\ee
and demanded that it commuted with the generator of gauge transformations
\be
G\equiv \int d^3 x \,f(x)\,\left(\pd_i E^i+ q\bar{\pi}\phi\right)
\ee
Using 
\bea
&\left[A^i,G\right]= -\pd^i f\nonumber\\
&\left[\phi,G\right]=-i q f \phi
\eea
The condition we need to fulfill is
\be
\left[\Phi_D,G\right]=-i \Phi_D\left(q f-\int \pd_{i^\prime} c_i(x,\xp) f(\xp) d^3 \xp\right)
\ee
That is
\be
\pd_{i^\prime}\pd^i c(x,\xp) = c(x,\xp) \d(x-\xp)
\ee
There are many solutions of this. One of the simplest is
\be
C(x)=i q\int d^3 \xp\,{(x-\xp)^i\over 4\pi |\vx-\vxp|^3}\, A_i(t,\vxp)
\ee

This operator creates a Coulomb field at time t
\be
\left[E^i(x),\Phi_D(\xp)\right]_{t=t^\prime}=-{q\over 4\pi}{(x-\xp)^i\over |\vx-\vxp|^3}\,\Phi_D(x)
\ee
This dressing can be generalized in various ways. One of the simplest is to introduce a Dirac string (Dirac called it a {\em Faraday line of force}). That is,
\be
C=-q \int_{-\infty}^x \,A(\xp_i) d\xp_i
\ee 
where the path of integration extends from infinity to the location of the charge at the point $x$. In this case the electric field is concentrated in the Dirac string.

\par
In \cite{Donnelly} Donnelly and Giddings have generalized this construction to the gravitational field, albeit to first order in the coupling constant, $\kappa$. For a scalar field the analogous dressed field is
\be
\Phi(x)\equiv \phi\left(x^\m+V^\m(x)\right)=\phi(x)+V^\m\pd_\m\phi+O(V^2)
\ee
where
\be
\d V^\m=-\kappa \xi^\m(x)
\ee
We can find vectors such that
\be
V^\m(x)=\kappa\int d^4 \xp f^{\m\n\l}(x,\xp)\,h_{\n\l}(\xp)
\ee
In order for this definition to be consistent with linearized diffeomorphism ($LDiff$)
\be
\d h_{\a\b}=\pd_\a\xi_\b+\pd_\b\xi_\a+O(\kappa)
\ee
\be
\pd_{\lp} f^{\m\n\l}={1\over 2}\,\d^4(x-\xp) \eta^{\m\n}
\ee
Preceding along these lines, Giddings \cite{Giddings} finds an obstruction to commutativity of gauge invariant observables associated to different regions of spacetime. Then locality fails for such dressed operators. Calculations are done using standard Dirac commutators stemming from the canonical theory.

\subsection{Subsystems.}
From the Haag algebraic  QFT theory, subalgebras of observables may be associated to regions of spacetime, and the subalgebra structure mirrors the topology of the spacetime manifold.
\par
A natural idea is then in the {\em quantum first} approach, to try to derive the spacetime structure from the net structure of the subalgebras of the von Neumann algebra of operators.
\par
We have just seen the difficulties for doing so in the presence of gravity, at least if the low energy limit os to be reached in a smooth way.
\par
Difficulties in the definition of subsystems stem from a couple of reasons: one is technical namely the so called type $III_\l$ property of the adequate subalgebras (which do not contain projectors at all), and the second is the existence of long range (gauge) fields.
\par
In QED it is possible to define \cite{Giddingss} localized uncharged observables, such as
\be
{\cal D}(x,\xp)\equiv \phi(x)\,e^{i q \int_{z_x}^{z_{\xp}}\,dz A(z)}\,\phi^*(\xp)
\ee
which creates a nontrivial field only in the string from the point $x$ to the point $\xp$. Then it commutes with all observables spacelike separated from this region. Interactions modify this property however. This string decays into a dipole field that extends to infinity. 
\par
In the presence of gravity, moreover, all localized excitations carry energy in such a way that gravity cannot be screened, so there is not even the analog of the  QED uncharged operator.
\par
In the absence of any hint from experiment it is not then clear how far can we go in this {\em quantum first} viewpoint.
\newpage
\section{Conclusions}
It should be hopefully clear from the preceding discussions that there are no definite conclusions as yet. Almost all avenues are still open. 
\par
Even the most straightforward idea of quantizing the metric tensor is not completely excluded, and some of the  most relevant work in the field is done by exploring just ordinary quantum field theory (QFT) amplitudes. It is curious that  Veltman's old idea that diagrams are more fundamental than the lagrangians themselves is now revamped in the work of Arkani-Hamed \cite{Arkani} and coworkers.

It is nevertheless clear that in QFT there is a rich non-perturbative sector, of which essentially nothing is known, in spite of the attempts of a great number of brilliant physicists in the second half of last century.
\par
In the particular case of  supersymmetric gauge theories there are some set of dualities \cite{Seiberg} (many of them inspired or even implied by string theory), but we are shy of understanding the detailed mechanism at work in the strong coupling sector. It should be said, however, that the only clear image \cite{SW} of confinement we have (due to Seiberg and Witten) is precisely in theories  with several (more than eight) supersymmetric charges.
\par
It is even conceivable that QFT contains string theory in some sense. The quantum field theoretical framework seems nowadays much richer than was formerly believed to be the case. In the non-supersymmetric case, it is true that most of our information on the non-perturbative sector in quantum chromodynamics (QCD) comes from the plethora of low energy data in hadronic physics. One wonders what would be our image of the infrared (IR) limit of QCD if we were not aware of this experimental information.
\par
 Quantum general relativity (quantizing the metric) is in some sense a gauge theory, but it is also a quite special one. Probably more complicated. It seems that it will be very difficult to understand precisely how a black hole is made out of self-interacting gravitons (or whatever elementary quanta are appropiate) as a bound state or confined state of sorts, before understanding the presumably simpler problem of how a glueball is made out of gluons in a Yang Mills theory such as QCD.
\par
Nevertheless, even in the absence of any experimental information, we cannot avoid to keep thinking on the relationship between two of the most successful theories in physics, namely General Relativity  and Quantum Mechanics, in whatever common ground they might share.
\par
 We are well aware that the chances of success are very slim, but, like in the famous poem {\em Ithaka} by Kavafis, the journey is fascinating, and we hace discovered many interesting vistas in our way. This is particularly true for string theory. Even if the main purpose of the theory (a unified theory of all interactions, including gravity) is not fulfilled, it cannot be denied that around research on this field many topics have flourished. In pure mathematics, of course, but not only. For example most of the modern techniques for doing advanced computations of amplitudes in gauge theories, associated to the names of Zvi Bern, Lance Dixon, David Kosower, and many others stem from string theory. For example, and against superficial appearances, to concentrate in on shell amplitudes and cubic vertices has proved enormously fruitful.
 \par
  But in spite of all that, Ithaka is still far, far away.

\section{Acknowledgements}
I am very grateful for discussions with  Jes\'us Anero, Jos\'e Gracia-Bond\'{\i}a  and Raquel Santos-Garc\'{\i}a, as well as to Steven Giddings, Pierre vanhove and Dejan Stojkovic for e-mail correspondence. I am grateful to the {\em Escuela de F\'{\i}sica}  of the University of Costa Rica for the invitation to lecture there on this topic. I can not forget the warm hospitality of Jos\'e Gracia-Bond\'{\i}a and Alejandro Jenkins, as well as of the  other faculty members in San Jos\'e in difficult times. This work has received funding from the Spanish Research Agency (Agencia Estatal de Investigacion) through the grant IFT Centro de Excelencia Severo Ochoa SEV-2016-0597, and the European Union's Horizon 2020 research and innovation programme under the Marie Sklodowska-Curie grants agreement No 674896 and No 690575. We also have been partially supported by FPA2016-78645-P(Spain).


\begin{thebibliography}{99}


\bibitem{Abbott}
  L.~F.~Abbott,
  `The Background Field Method Beyond One Loop,''
  Nucl.\ Phys.\ B {\bf 185} (1981) 189.
  doi:10.1016/0550-3213(81)90371-0



\bibitem{firewalls}
  A.~Almheiri, D.~Marolf, J.~Polchinski and J.~Sully,
  ``Black Holes: Complementarity or Firewalls?,''
  JHEP {\bf 1302} (2013) 062
  doi:10.1007/JHEP02(2013)062
  [arXiv:1207.3123 [hep-th]].
\bibitem{AlvarezGaume}
L.~Alvarez-Gaume and M.~Vazquez-Mozo,
``Topics in string theory and quantum gravity,''
[arXiv:hep-th/9212006 [hep-th]].
%
\bibitem{Alvarez}
E.~Alvarez,
``Quantum Gravity: A Pedagogical Introduction To Some Recent Results,''
Rev.\ Mod.\ Phys.\  {\bf 61} (1989) 561.\\
``Some general problems in quantum gravity,''
CERN-TH-6257-91.


\bibitem{Alvarezz}
E. Alvarez, ``Can one tell Einstein's unimodular theory from Einstein's general
  JHEP {\bf 0503} (2005) 002
  [arXiv:hep-th/0501146].\\




\bibitem{Anderson}
  J.~L.~Anderson and P.~G.~Bergmann,
  ``Constraints in covariant field theories,''
  Phys.\ Rev.\  {\bf 83} (1951) 1018.
  doi:10.1103/PhysRev.83.1018


\bibitem{Anselmi}
  D.~Anselmi,
  ``Renormalization and causality violations in classical gravity coupled with quantum matter,''
  JHEP {\bf 0701} (2007) 062
  doi:10.1088/1126-6708/2007/01/062
  [hep-th/0605205].

  
\bibitem{Arkani}
  N.~Arkani-Hamed and J.~Trnka,
  ``The Amplituhedron,''
  JHEP {\bf 1410} (2014) 030
  doi:10.1007/JHEP10(2014)030
  [arXiv:1312.2007 [hep-th]].


\bibitem{ADM}R. Arnowitt, S. Deser, and C. W. Misner,
`` Canonical Variables for General Relativity''
Phys. Rev. 117, 1595-1602 (1960)  
 Hep :: HepNames :: Institutions :: Conferences :: Jobs :: Experiments :: Journals :: Help

\bibitem{Ashtekar}
  A.~Ashtekar,
  ``New Variables for Classical and Quantum Gravity,''
  Phys.\ Rev.\ Lett.\  {\bf 57} (1986) 2244.
  doi:10.1103/PhysRevLett.57.2244


\bibitem{Banks}
T.~Banks,
``T C P, Quantum Gravity, the Cosmological Constant and All That...,''
Nucl. Phys. B \textbf{249} (1985), 332-360
doi:10.1016/0550-3213(85)90020-3

\bibitem{Bekenstein}
J. Bekenstein, {\em Black  Holes And Entropy}
Phys.Rev. D7 (1973) 2333\\
(with V. Mukhanov), ``Spectroscopy Of The Quantum Black Hole'', Commun.Math.Phys. 125 (1989) 417


\bibitem{Bern}
Z.~Bern,
``Perturbative quantum gravity and its relation to gauge theory,''
Living Rev.\ Rel.\  {\bf 5} (2002) 5
[arXiv:gr-qc/0206071].\\
  Z.~Bern, J.~J.~Carrasco, M.~Chiodaroli, H.~Johansson and R.~Roiban,
  ``The Duality Between Color and Kinematics and its Applications,''
  arXiv:1909.01358 [hep-th].\\
  Z.~Bern, C.~Cheung, R.~Roiban, C.~H.~Shen, M.~P.~Solon and M.~Zeng,
  ``Black Hole Binary Dynamics from the Double Copy and Effective Theory,''
  JHEP {\bf 1910} (2019) 206
  doi:10.1007/JHEP10(2019)206
  [arXiv:1908.01493 [hep-th]].



\bibitem{BjerrumBohr}
  N.~E.~Bjerrum-Bohr, J.~F.~Donoghue, B.~R.~Holstein,
  ``Quantum gravitational corrections to the nonrelativistic scattering potential of two masses,''
  Phys.\ Rev.\  {\bf D67}, 084033 (2003).
  [hep-th/0211072].



\bibitem{Bondi}
  H.~Bondi, M.~G.~J.~van der Burg and A.~W.~K.~Metzner,
  ``Gravitational waves in general relativity. 7. Waves from axisymmetric isolated systems,''
  Proc.\ Roy.\ Soc.\ Lond.\ A {\bf 269} (1962) 21.
  doi:10.1098/rspa.1962.0161\\
  
\bibitem{Books}
R.P. Feynman, 
{\em Lectures on Gravitation},
                 (University of Bangalore Press,1997)\\
 B.~S.~DeWitt,
  ``The global approach to quantum field theory. Vol. 1, 2,''
  Int.\ Ser.\ Monogr.\ Phys.\  {\bf 114} (2003) 1.\\
Claus Kiefer,
"Quantum Gravity"
(Clarendon Press, Oxford, 2004)\\
Carlo Rovelli,
"Quantum gravity"
(Cambrige UP,2004)\\
 Viatcheslav Mukhanov, Sergei Winitzki 
 "Introduction to Quantum Effects in Gravity"
 (Cambridge UP,2007)
 


\bibitem{Brown}
  J.~D.~Brown, S.~R.~Lau and J.~W.~York, Jr.,
  ``Action and energy of the gravitational field,''
  gr-qc/0010024.
  


\bibitem{Carlip}
  S.~Carlip,
  ``Is Quantum Gravity Necessary?,''
  Class.\ Quant.\ Grav.\  {\bf 25} (2008) 154010
  doi:10.1088/0264-9381/25/15/154010
  [arXiv:0803.3456 [gr-qc]].
%
\bibitem{Chang}
  C.~C.~Chang, J.~M.~Nester and C.~M.~Chen,
  ``Pseudotensors and quasilocal gravitational energy momentum,''
  Phys.\ Rev.\ Lett.\  {\bf 83} (1999) 1897
  doi:10.1103/PhysRevLett.83.1897
  [gr-qc/9809040].
 
\bibitem{Dai}
 D.~C.~Dai and D.~Stojkovic,
 ``Inconsistencies in Verlinde's emergent gravity,''
 JHEP {\bf 1711}, 007 (2017)
 doi:10.1007/JHEP11(2017)007
 [arXiv:1710.00946 [gr-qc]].

 
\bibitem{DeWitt}
  B.~S.~DeWitt,
  ``Dynamical theory of groups and fields,''
  Conf.\ Proc.\ C {\bf 630701} (1964) 585
   [Les Houches Lect.\ Notes {\bf 13} (1964) 585].\\
``Quantum Theory Of Gravity. 1. The Canonical Theory,''
Phys.\ Rev.\  {\bf 160} (1967) 1113.\\
``Quantum Theory of Gravity. II. The Manifestly Covariant Theory'',
                          Phys. Rev. 162, 1195-1239 (1967).

\bibitem{Dirac}
PAM Dirac,
"Lectures in quantum mechanics"
(Yeshiva, NY, 1964)\\
  A.~J.~Hanson, T.~Regge and C.~Teitelboim,
  ``Constrained Hamiltonian Systems,''
  RX-748, PRINT-75-0141 (IAS,PRINCETON).



\bibitem{Diracc}
  P.~A.~M.~Dirac,
  ``Gauge invariant formulation of quantum electrodynamics,''
  Can.\ J.\ Phys.\  {\bf 33} (1955) 650.
  doi:10.1139/p55-081


\bibitem{Donnelly}
  W.~Donnelly and S.~B.~Giddings,
  ``Diffeomorphism-invariant observables and their nonlocal algebra,''
  Phys.\ Rev.\ D {\bf 93} (2016) no.2,  024030
   Erratum: [Phys.\ Rev.\ D {\bf 94} (2016) no.2,  029903]
  doi:10.1103/PhysRevD.94.029903, 10.1103/PhysRevD.93.024030
  [arXiv:1507.07921 [hep-th]].
                   

\bibitem{Donoghue}
J.~F.~Donoghue,
``General Relativity As An Effective Field Theory: The Leading Quantum
Corrections,''
Phys.\ Rev.\ D {\bf 50} (1994) 3874
[arXiv:gr-qc/9405057].



\bibitem{Duff}
M.~J.~Duff,
``Inconsistency Of Quantum Field Theory In Curved Space-Time,''
ICTP/79-80/38.
{\it Talk presented at 2nd Oxford Conf. on Quantum Gravity, Oxford, Eng., Apr 1980}\\
  ``Quantum Tree Graphs and the Schwarzschild Solution,''
  Phys.\ Rev.\  D {\bf 7} (1973) 2317.



\bibitem{ER}
  A.~Einstein and N.~Rosen,
  ``The Particle Problem in the General Theory of Relativity,''
  Phys.\ Rev.\  {\bf 48} (1935) 73.
  doi:10.1103/PhysRev.48.73


\bibitem{EPR}
  A.~Einstein, B.~Podolsky and N.~Rosen,
  ``Can quantum mechanical description of physical reality be considered complete?,''
  Phys.\ Rev.\  {\bf 47} (1935) 777.
  doi:10.1103/PhysRev.47.777

\bibitem{E19}
[ A. Einstein, 
Siz. Preuss. Acad. Scis. 1919,
 english translation in
 " The principle of relativity"
  (A. Einstein et al. eds., Dover).



\bibitem{Eisenhart}
L.P. Eisenhart,
"Riemannian geometry"
(Princetin,NJ)

                   



\bibitem{Fock}
Vladimir A. Fock,
"Theory of Space, Time and Gravitation",
(Pergamon press, 1963)



\bibitem{Friedan}
D.~Friedan,
``Nonlinear Models In Two Epsilon Dimensions,''
Phys.\ Rev.\ Lett.\  { 45} (1980) 1057.



\bibitem{GH}
  G.~W.~Gibbons and S.~W.~Hawking,
  ``Cosmological Event Horizons, Thermodynamics, and Particle Creation,''
  Phys.\ Rev.\ D {\bf 15} (1977) 2738.
  doi:10.1103/PhysRevD.15.2738



\bibitem{Giddings}
  S.~B.~Giddings,
  ``Hilbert space structure in quantum gravity: an algebraic perspective,''
  JHEP {\bf 1512} (2015) 099
  doi:10.1007/JHEP12(2015)099
  [arXiv:1503.08207 [hep-th]].\\
  ``Quantum gravity: a quantum-first approach,''
  LHEP {\bf 1} (2018) no.3,  1
  doi:10.31526/LHEP.3.2018.01
  [arXiv:1805.06900 [hep-th]].\\
  "Quantum-first gravity",
  arXiv:1803.04973\\
"Gravitational dressing, soft charges, and perturbative gravitational splitting",
arXiv:1903.06160
.
  \bibitem{Giddingss}
S.~B.~Giddings, D.~Marolf and J.~B.~Hartle,
  ``Observables in effective gravity,''
  Phys.\ Rev.\  D {\bf 74} (2006) 064018
  [arXiv:hep-th/0512200].


\bibitem{Gliozzi}
F.~Gliozzi, J.~Scherk and D.~I.~Olive,
``Supersymmetry, Supergravity Theories And The Dual Spinor Model,''
Nucl.\ Phys.\ B { 122} (1977) 253.


\bibitem{Goroff}
M.~H.~Goroff and A.~Sagnotti,
``The Ultraviolet Behavior Of Einstein Gravity,''
Nucl.\ Phys.\ B {\bf 266}, 709 (1986).

\bibitem{GraciaBondia}
J.~M.~Gracia-Bondia,
``Notes on 'quantum gravity' and non-commutative geometry,''
Lect. Notes Phys. 807 (2010), 3-58
doi:10.1007/978-3-642-11897-5-1
[arXiv:1005.1174 [hep-th]].

\bibitem{Green}
M.~B.~Green and J.~H.~Schwarz,
``Anomaly Cancellation In Supersymmetric D=10 Gauge Theory And Superstring Theory,''
Phys.\ Lett.\ B { 149} (1984) 117.\\
M.~B.~Green, J.~H.~Schwarz and E.~Witten,
``Superstring Theory. Vol. 1: Introduction,''
\\
``Superstring Theory. Vol. 2: Loop Amplitudes, Anomalies And Phenomenology,''

\bibitem{GRV}
M.~B.~Green, J.~G.~Russo and P.~Vanhove,
``Ultraviolet properties of maximal supergravity,''
Phys. Rev. Lett. \textbf{98} (2007), 131602
doi:10.1103/PhysRevLett.98.131602
[arXiv:hep-th/0611273 [hep-th]].
\cite{GreenBj}
\bibitem{Bjornsson:2010wm}
J.~Bjornsson and M.~B.~Green,
``5 loops in 24/5 dimensions,''
JHEP \textbf{08} (2010), 132
doi:10.1007/JHEP08(2010)132
[arXiv:1004.2692 [hep-th]].
\bibitem{Hartle}
J.~B.~Hartle,
``Space-time quantum mechanics and the quantum mechanics of space-time,''
arXiv:gr-qc/9304006.\\
  J.~B.~Hartle and S.~W.~Hawking,
  ``Wave Function of the Universe,''
  Phys.\ Rev.\ D {\bf 28} (1983) 2960
   [Adv.\ Ser.\ Astrophys.\ Cosmol.\  {\bf 3} (1987) 174].
  doi:10.1103/PhysRevD.28.2960
%

\bibitem{Hawking}
S. Hawking, {\em  Particle Creation By Black Holes}
 Commun.Math.Phys. 43 (1975) 199





 
\bibitem{'tHooft}
G.~'t Hooft,
``Dimensional Reduction In Quantum Gravity,''
arXiv:gr-qc/9310026.\\
``The Scattering matrix approach for the quantum black hole: An Overview,''
  Int.\ J.\ Mod.\ Phys.\  {\bf A11}, 4623-4688 (1996).
  [gr-qc/9607022].\\
  ``Quantum gravity without space-time singularities or horizons,''
  arXiv:0909.3426 [gr-qc].


\bibitem{thv}
G.~'t Hooft and M.~J.~Veltman,
``One Loop Divergencies In The Theory Of Gravitation,''
Annales Poincare Phys.\ Theor.\ A { 20} (1974) 69.




\bibitem{Hull}
C.~M.~Hull and P.~K.~Townsend,
``Unity of superstring dualities,''
Nucl.\ Phys.\ B { 438} (1995) 109
[arXiv:hep-th/9410167].




\bibitem{Israel} 
  W.~Israel,
  ``Thermo field dynamics of black holes,''
  Phys.\ Lett.\ A {\bf 57}, 107 (1976).
  doi:10.1016/0375-9601(76)90178-X


\bibitem{Jacobson}
T.~Jacobson,
``Thermodynamics of space-time: The Einstein equation of state,''
Phys. Rev. Lett. \textbf{75} (1995), 1260-1263
doi:10.1103/PhysRevLett.75.1260
[arXiv:gr-qc/9504004 [gr-qc]].




\bibitem{Kibble}
  T.~W.~B.~Kibble and S.~Randjbar-Daemi,
  ``Nonlinear Coupling of Quantum Theory and Classical Gravity,''
  J.\ Phys.\ A {\bf 13} (1980) 141.
  doi:10.1088/0305-4470/13/1/015

%

\bibitem{Makeenko}
  Y.~Makeenko,
  ``Methods of contemporary gauge theory,''
  doi:10.1017/CBO9780511535147\\
Y.~Makeenko and A.~A.~Migdal,
``Exact Equation for the Loop Average in Multicolor QCD,''
Phys. Lett. B \textbf{88} (1979), 135
doi:10.1016/0370-2693(79)90131-X


\bibitem{Maldacena}
J. Maldacena, 
                   {\em The large N limit of superconformal field
                    theories and supergravity},
                   Adv.Theor.Math.Phys.2:231-252,1998, 
                    {\tt hep-th/9711200}. 





\bibitem{Milnor}
  J.~Milnor,
  ``Remarks On Infinite Dimensional Lie Groups,''
{\it  In *Les Houches 1983, Proceedings, Relativity, Groups and Topology, Ii*, 1007-1057}

\bibitem{Moller}
C.~Moller,
``Further remarks on the localization of the energy in the general theory of relativity,''
Annals Phys. \textbf{12} (1961), 118-133
doi:10.1016/0003-4916(61)90148-8


\bibitem{Montonen}
C.~Montonen and D.~I.~Olive,
``Magnetic Monopoles as Gauge Particles?,''
Phys. Lett. B \textbf{72} (1977), 117-120
doi:10.1016/0370-2693(77)90076-4


\bibitem{Myers}
R.~C.~Myers and M.~Pospelov,
``Experimental challenges for quantum gravity,''
arXiv:hep-ph/0301124.

\bibitem{Tomas}
Tomas Ortin,
"Gravity and strings"
(Cambridge UP,2004)\\
Luis Iba\~nez and Angel Uranga,
"String theory and particle physics
(Cambridge UP, 2012)

\bibitem{Osborn}
H.~Osborn,
``Topological Charges for N=4 Supersymmetric Gauge Theories and Monopoles of Spin 1,''
Phys. Lett. B \textbf{83} (1979), 321-326
doi:10.1016/0370-2693(79)91118-3

\bibitem{Pioline}
B.~Pioline,
``String theory integrands and supergravity divergences,''
JHEP \textbf{02} (2019), 148
doi:10.1007/JHEP02(2019)148
[arXiv:1810.11343 [hep-th]].\\
M.~B.~Green, S.~D.~Miller, J.~G.~Russo and P.~Vanhove,
`Eisenstein series for higher-rank groups and string theory amplitudes,''
Commun. Num. Theor. Phys. \textbf{4} (2010), 551-596
doi:10.4310/CNTP.2010.v4.n3.a2
[arXiv:1004.0163 [hep-th]].\\
M.~B.~Green, J.~G.~Russo and P.~Vanhove,
``String theory dualities and supergravity divergences,''
JHEP \textbf{06} (2010), 075
doi:10.1007/JHEP06(2010)075
[arXiv:1002.3805 [hep-th]].
\bibitem{Polchinski}
  J.~Polchinski,
  ``What is string theory?,''
  hep-th/9411028.


\bibitem{Poisson}
  E.~Poisson,
  ``A Relativist's Toolkit: The Mathematics of Black-Hole Mechanics,''
  doi:10.1017/CBO9780511606601

\bibitem{PolchinskiD}
J.~Polchinski,
``Dirichlet-Branes and Ramond-Ramond Charges,''
Phys.\ Rev.\ Lett.\  { 75} (1995) 4724
[arXiv:hep-th/9510017].

\bibitem{Polchinskis}
J.~Polchinski,
``String Theory. Vol. 1: An Introduction To The Bosonic String,''
\\
``String Theory. Vol. 2: Superstring Theory And Beyond,''



\bibitem{Ryu}
  S.~Ryu and T.~Takayanagi,
  ``Holographic derivation of entanglement entropy from AdS/CFT,''
  Phys.\ Rev.\ Lett.\  {\bf 96} (2006) 181602
  doi:10.1103/PhysRevLett.96.181602
  [hep-th/0603001].
  
 
 
\bibitem{Seiberg}
  N.~Seiberg,
  `Electric - magnetic duality in supersymmetric nonAbelian gauge theories,''
  Nucl.\ Phys.\ B {\bf 435} (1995) 129
  doi:10.1016/0550-3213(94)00023-8
  [hep-th/9411149].


\bibitem{Strominger}
A.~Strominger and C.~Vafa,
``Microscopic Origin of the Bekenstein-Hawking Entropy,''
Phys.\ Lett.\ B { 379} (1996) 99
[arXiv:hep-th/9601029].
\bibitem{SW}
  N.~Seiberg and E.~Witten,
  ``Electric - magnetic duality, monopole condensation, and confinement in N=2 supersymmetric Yang-Mills theory,''
  Nucl.\ Phys.\ B {\bf 426} (1994) 19
   Erratum: [Nucl.\ Phys.\ B {\bf 430} (1994) 485]
  doi:10.1016/0550-3213(94)90124-4, 10.1016/0550-3213(94)00449-8
  [hep-th/9407087].


\bibitem{Susskind}
L.~Susskind,
``The World as a hologram,''
J.\ Math.\ Phys.\  { 36} (1995) 6377
[arXiv:hep-th/9409089].

 
\bibitem{Takahashi}
Y.~Takahashi and H.~Umezawa,
``Thermo field dynamics,''
Int. J. Mod. Phys. B \textbf{10} (1996), 1755-1805
doi:10.1142/S0217979296000817


\bibitem{Thiemann}
  T.~Thiemann,
  `Modern Canonical Quantum General Relativity,''
  doi:10.1017/CBO9780511755682


\bibitem{Unruh}
W.~Unruh,
``Notes on black hole evaporation,''
Phys. Rev. D \textbf{14} (1976), 870
doi:10.1103/PhysRevD.14.870

 \bibitem{unimodular}
  W.~G.~Unruh,
  ``A Unimodular Theory of Canonical Quantum Gravity,''
  Phys.\ Rev.\ D {\bf 40} (1989) 1048.
  doi:10.1103/PhysRevD.40.1048

\bibitem{VanRaamsdonk}
  M.~Van Raamsdonk,
  ``Building up spacetime with quantum entanglement,''
  Gen.\ Rel.\ Grav.\  {\bf 42} (2010) 2323
   [Int.\ J.\ Mod.\ Phys.\ D {\bf 19} (2010) 2429]
  doi:10.1007/s10714-010-1034-0, 10.1142/S0218271810018529
  [arXiv:1005.3035 [hep-th]].


\bibitem{Verlinde}
E.~P.~Verlinde,
``On the Origin of Gravity and the Laws of Newton,''
JHEP \textbf{04} (2011), 029
doi:10.1007/JHEP04(2011)029
[arXiv:1001.0785 [hep-th]].\\
``Emergent Gravity and the Dark Universe,''
SciPost Phys. \textbf{2} (2017) no.3, 016
doi:10.21468/SciPostPhys.2.3.016
[arXiv:1611.02269 [hep-th]].

\bibitem{Vilenkin}
  A.~Vilenkin,
  ``The Birth of Inflationary Universes,''
  Phys.\ Rev.\ D {\bf 27} (1983) 2848.
  doi:10.1103/PhysRevD.27.2848

\bibitem{Weinberg}
 S. Weinberg,{\em Ultraviolet divergences in quantum theories of gravitation},
                     in {\em General relativity: an Einstein centenary volume}, 
                     S.W. Hawking and W. Israel eds.
                     (Cambridge University press,1979).\\
                     ``Effective Field Theory, Past and Future,''
  arXiv:0908.1964 [hep-th].
\bibitem{Weinbergqm}
  S.~Weinberg,
  ``Testing Quantum Mechanics,''
  Annals Phys.\  {\bf 194} (1989) 336.
  doi:10.1016/0003-4916(89)90276-5


\bibitem{Wheeler}
J.A. Wheeler,
"Infomation, physics, quantum:the search for links",
Proceedings III Int. Simp. on foundations of quantum mechanics,
(Tokyo 1989)

\bibitem{Witten}
 E.~Witten,
                 {\sl Anti de Sitter Space and Holography},
                 {\tt hep-th/9802150};\\
 {\sl Anti-de Sitter space,
                   thermal phase transition and confinement in gauge
                   theories}
                 {\tt hep-th/9803131}.
                
\bibitem{Witten3}
E.~Witten,
``(2+1)-Dimensional Gravity As An Exactly Soluble System,''
Nucl.\ Phys.\ B { 311} (1988) 46.



\bibitem{Woodard}
  R.~P.~Woodard,
  ``Avoiding dark energy with 1/r modifications of gravity,''
  Lect.\ Notes Phys.\  {\bf 720} (2007) 403
  [astro-ph/0601672].
































\end{thebibliography}
\end{document}